\documentclass{article}
\usepackage[utf8]{inputenc}
\usepackage{amssymb}
\usepackage[overload]{empheq}
\usepackage{braket}
\usepackage[english]{babel}
\usepackage{amsmath}
\usepackage{xcolor}
\usepackage[normalem]{ulem}

\usepackage[final]{pdfpages}
\usepackage{csquotes}
\usepackage[T1]{fontenc}
\usepackage{nomencl}
\usepackage{hyperref}
\usepackage{multicol}

\usepackage[linesnumbered,ruled,vlined]{algorithm2e}

\SetCommentSty{mycommfont}

\SetKwInput{KwInput}{Input}                
\SetKwInput{KwOutput}{Output}  

\title{Echo chambers in the Ising model and implications on the mean magnetization}
\author{Talia Baravi, Ofer Feinerman, Oren Raz }
\date{\today}

\begin{document}
\maketitle
\begin{abstract}
    The echo-chamber effect is a common term in opinion dynamic modeling to describe how a person's opinion might be artificially enhanced as it is reflected back at her through social interactions. Here, we study the existence of this effect in statistical mechanics models, which are commonly used to study opinion dynamics. We show that the Ising model does not exhibit echo-chambers, but this result is a consequence of a special symmetry. We then distinguish between three types of models: (i) those with a strong echo-chamber symmetry, that have no echo-chambers at all; (ii) those with a weak echo-chamber symmetry that can exhibit echo-chambers but only if there are external fields in the system, and (iii)  models without echo-chamber symmetry that generically have echo-chambers. We use these results to construct an efficient algorithm to efficiently and precisely calculate magnetization in arbitrary tree networks. Finally, We apply this algorithm to study two systems: phase transitions in the random field Ising model on a Bethe lattice and the influence optimization problem in social networks.     
\end{abstract}

\section{Introduction}

The formation of echo-chambers in social media \cite{jamieson2008echo,currin2021depolarization}  is thought to strongly affect how opinions spread across a social network and contribute to increased social polarity. This occurs as like-minded people cluster in disjoint networks wherein information produced by any individual is reflected back at her \cite{cinelli2021echo} to increase her confidence artificially. This effect is countered, to a degree, by the diversity of opinions in a network which, naturally, exceeds that of any single individual \cite{dubois2018echo,bruns2017australian,del2018echo}.
Indeed, the formation of echo-chambers appears to depend on the specific interaction rules adopted by different platforms \cite{cinelli2021echo,Huszare2025334119}. 

In this paper, we explore the existence of echo-chambers in statistical mechanics models that are often used to study opinion and influence dynamics but are common in many additional contexts -- from magnetic systems to neural networks. We focus most of our attention on the famous Ising model, which describes a ferromagnet as a network of coupled spins \cite{pathria2011statistical}. This model was originally developed to describe equilibrium magnetic systems in statistical mechanics, but is often used to explore how interactions between the individual entities reflect on large-scale system properties, for example, in the context of influence dynamics \cite{mobilia2007role,hartnett2016heterogeneous,liu2010influence,kempe2003maximizing,galam2000universality}. A natural question is therefore whether echo chambers exist in this model. We would say that a spin experiences an \emph{ echo chamber} effect if, when exposed to an external local field, its mean magnetization depends on whether it is isolated or linked to a neutral network. Here, a \emph{neutral network} with respect to the specific spin means a network that does not bias the mean magnetization of this spin when no external field is applied on it. In other words, the spin echo-chamber effect occurs when the connection with the network does not bias an unbiased spin but amplifies the influence of any external field applied on the spin.

Intuitively, one might expect the Ising model to display spin echo-chambers: Consider a spin connected to a network with no external fields on any of its spins and is, therefore, a neutral network. An external field on the specific spin connected to the network biases it in a certain direction and produces mean magnetization. Once this spin is connected to the network, this magnetization breaks the $\pm$ symmetry and biases the neighboring spins in the same direction. The network then has non-zero magnetization. One might expect it would feedback on the local spin to change its bias and mean magnetization, generating an echo-chamber in the system. Surprisingly, we find that although it might exist in other models, this effect is absent from the Ising model. We then identify three cases that can occur in different models: (i) Strong echo-chamber symmetry -- models in which echo-chambers never appear in a neutral network. The Ising model belongs to this class; (ii) Weak echo-chamber symmetry -- where there are no echo chambers in networks that do not have external fields other than on the specific spin, but there might be echo-chambers in networks that have additional biasing fields that balance each other on the specific spin, generating a neutral network. The XY, the vector Potts, and the Heisenberg models belong to this class; (iii) Models without any echo-chamber symmetry, which generically include the echo-chamber effect. For example, the spin $1$ Blume-Capel model belongs to this class.

To prove the absence of spin echo chambers in the Ising model, we use a simple but powerful tool -- the \emph{effective field method}. The mean magnetization of a spin is affected by the external field applied to it and its interactions with the other spins. We show that the two effects decouple and that the interactions with the rest of the network can be fully described by an \emph{effective field}. While an effective field could always be defined for any model, in the Ising model, it is independent of the external field applied on the spin itself. This property turns out to prevent the spin echo chamber effect. Loosely speaking, an effective field implies the lack of echo chambers even in non-neutral networks.

The implications of the effective field method go beyond the proof that some models do not exhibit a spin echo chamber effect. To demonstrate this, we applied it to construct an efficient computational algorithm for an exact calculation of the mean magnetization of each spin in tree networks with general interaction strengths, local external fields, and temperature. Although Monte-Carlo and similar methods are usually efficient in approximating the mean magnetization in such systems to high accuracy, exact calculations as in our algorithm are typically limited by the fact that the partition function grows exponentially with system size. As a proof of concept, we demonstrate this algorithm by calculating the quenched average of the random field Ising model on the Bethe lattice \cite{bruinsma1984random,nowotny2001phase,bleher1998phase}. This system is known to have a phase transition as a function of the temperature and the magnitude of the quenched disordered field. Our method enabled us to exactly calculate the mean magnetization in each realization of the random field and average these to directly demonstrate quenched averaged magnetization on a relatively large system.

Finally, we connect our findings back to the field of social influence networks \cite{friedkin1997social}. Specifically, we consider the optimal ways to exert influence on a network given limited resources \cite{yang2006mining,liu2010influence}. Using the analogy between social networks and the Ising model \cite{liu2010influence,stauffer2008social} it was recently shown that, to maximize influence, either hubs or leaf nodes should be targeted \cite{lynn2017statistical,romero2020continuous}. We validate and expand on these results on a large Barabasi-Albert network \cite{barabasi2014network}, and without the mean-field approximation, \cite{lynn2016maximizing} commonly used to analyze such systems.

\section{The basic model}
We start by considering the Ising spin model on a general undirected graph $G$ in which at each node $i$ there is a spin, denoted by $\sigma_i$. We assume an arbitrary symmetric interaction term $J_{ij}=J_{ji}$ between spins $\sigma_i$ and $\sigma_j$, if the nodes $i$ and $j$ are connected in the the graph $G$. Furthermore, we allow the external field $h_i$ to differ between nodes. The Hamiltonian of this model is given by:
\begin{equation}\label{Eq:Hamiltonian_Def}
    \mathcal{H}(\vec\sigma)=-\sum_{\{i,j\}\in G}J_{ij}\sigma_i\sigma_j-\sum_i h_i \sigma_i,
\end{equation} 
where $\vec\sigma = (\sigma_1,...,\sigma_N)$ is a specific configuration of the spins, $J_{ij}$ is the coupling strength between the spin $i$ and spin $j$, $h_i$ is the external field on the spin $i$, the first summation is over all the edges of the graph and the last summation is over all spins.

For a system described by the above Hamiltonian, the average magnetization of the $k^{th}$ spin in thermal equilibrium is given by:
\begin{equation}\label{Eq:MeanSigma_k}
    \braket{\sigma_k}=\frac{\sum_{\{\sigma\}}\sigma_k e^{\mathcal{-\beta H}(\vec\sigma)}}{\sum_{\{\sigma\}}e^{\mathcal{-\beta H}(\vec\sigma)}} = \frac{\partial_{h_k}Z}{\beta Z},
\end{equation}
where $\sum_{\{\sigma\}}$ is a summation over all spin configurations, $\beta=T^{-1}$ is the inverse temperature in units where the Boltzmann constant is set to 1, and 
\begin{eqnarray}
    Z = \sum_{\{\sigma\}}e^{\mathcal{-\beta H}(\vec\sigma)}
\end{eqnarray}
is the partition function associated with the model. For the specific case of a single spin, which is used in the definition of the spin echo-chamber effect, the partition function is
\begin{eqnarray}
    Z_1 = 2\cosh(\beta h),
\end{eqnarray}
and the mean magnetization is given by
\begin{eqnarray}\label{Eq:SingleSpinMean}
    \braket{\sigma} = \tanh(\beta h).
\end{eqnarray}

Even though we focus much of our work on this basic model to demonstrate our findings, some of our results hold for a more general class of models,  which we call ``echo-chamber symmetric'' models, as discussed in Sec.(\ref{Sec:OtherModel}).

\section{Illustrative examples in two spin systems}

To demonstrate the echo chamber effect, we start by looking at the simplest network graph, which includes two connected nodes. We show that the Ising model does not exhibit an echo chamber in this example, while a spin $1$ model does. In addition, we provide some heuristic intuition to these results. A more comprehensive analysis expands these simple examples in the following sections.

\subsection{The Ising model lacks an echo chamber}
Let us start by considering a single Ising spin, $\sigma_1$, in an external  field of $h_1=1$ and inverse temperature $\beta=1$. As stated above, the mean magnetization of this spin is simply :
\begin{eqnarray}
    \braket{\sigma_1} = \tanh(\beta h_1) = \tanh(1)\sim 0.76.
\end{eqnarray}

Next, we connect $\sigma_1$ to the simplest possible network by adding just one more spin, $\sigma_2$, coupled with $J_{12}=1$. There is no external field on $\sigma_2$, namely $h_2=0$. The Hamiltonian in this case, is simply

\begin{equation}\label{Eq:Hamiltonian_Def}
    \mathcal{H}(\vec\sigma)=-\sigma_1\sigma_2- \sigma_1.
\end{equation}

Let us calculate the mean magnetization of this second spin. Direct calculation shows that in this case: 
\begin{eqnarray}
    \braket{\sigma_2} =\frac{e^{2} + e^{-2} - 2}{e^{2} + e^{-2} + 2}\sim 0.58. 
\end{eqnarray}
Although no field is directly applied to $\sigma_2$, its coupling with $\sigma_1$ leads to a mean magnetization which agrees with the sign of the external field $h_1$. Since $\braket{\sigma_2}>0$, one might expect that this magnetization may feedback onto  $\sigma_1$ and increase its alignment with the field $h_1$. Indeed, if one replaces $\sigma_2$ with a fixed magnet of constant magnetization equal to $\braket{\sigma_2}$ then this would increase the mean magnetization of $\sigma_1$.

Directly calculating the mean magnetization of $\sigma_1$ in the coupled system we obtain:
\begin{eqnarray}
    \braket{\sigma_1} =\frac{e^{2} - e^{-2}}{e^{2} + e^{-2} + 2}\sim 0.76, 
\end{eqnarray}
which is exactly equal to its mean magnetization in the uncoupled system. Therefore, the expectation for positive feedback does not hold and this system does not exhibit an echo chamber.

The effect of  $\sigma_2$  on  $\sigma_1$ deviates from that of a constant magnet with the same mean magnetization. Therefore, it must be the fluctuations of $\sigma_2$ responsible for the absence of echo-chamber. Using the ergodicity of the system, which implies that the ensemble average is identical to the time average, we can interpret $\braket{\sigma_2}>0$ as if, on average, $\sigma_2$ spends time $t_+$ in the $\sigma_2=+1$ state and time $t_-$ in the $\sigma_2=-1$ state, such that
\begin{eqnarray}
\frac{t_+-t_-}{t_++t_-}=\braket{\sigma_2}.    
\end{eqnarray}

\begin{figure}[h]
  \centering
    \includegraphics[width=0.7 \textwidth]{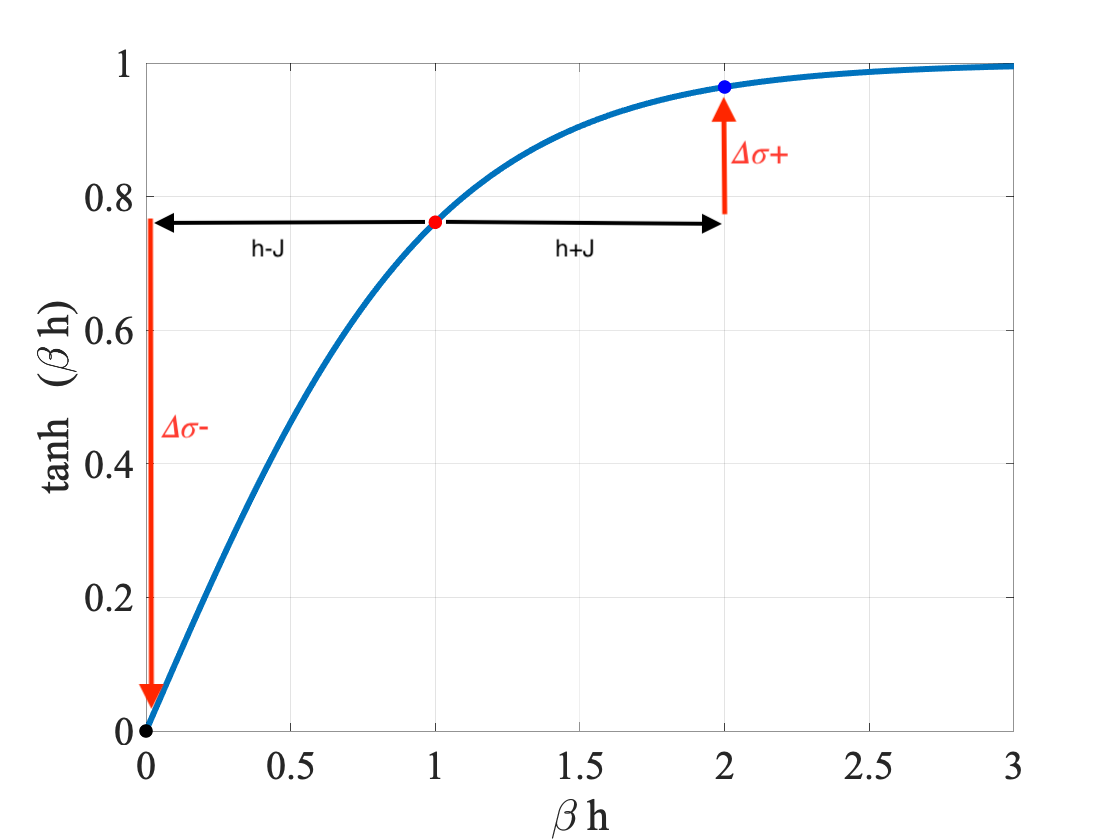}
\caption{The blue line is the function $\tanh(\beta h)$, which relates between average magnetization and external field in the Ising model. The red point is $\braket{\sigma_1}$ for an isolated spin with $\beta=h=1$. Coupling this spin to a second spin, $\sigma_2$, with coupling constant $J=1$, changes the external influence on $\sigma_1$ to $\beta(h+J)$ when $\sigma_2=+1$ and to $\beta(h-J)$ when $\sigma_2=-1$ (dark blue and black points, respectively). The concave nature of the hyperbolic tangent implies that for $h>0$ the $\sigma_2=-1$ cause a larger change in the mean magnetization compared to $\sigma_2=+1$ (compare the magnitudes of the red arrows). The time difference $\sigma_2$ spends on each of these states exactly cancels this bias so that the two spin Ising model,lacks an echo-chamber.}
\label{fig:HypTan}
\end{figure}

A key point in this intuitive explanation is then to note that because of the external field $h_1$,  $\sigma_2$ affects the mean magnetization of $\sigma_1$ stronger when $\sigma_2=-1$ in comparison to the case $\sigma_2=+1$. This is a result of the concave nature of Eq.(\ref{Eq:SingleSpinMean}) for $h>0$ (see Fig. \ref{fig:HypTan}): Since $h_1>0$, a bias towards $h_1+J$ when $\sigma_2=+1$ has a smaller effect on $\braket{\sigma_1}$ in comparison to the bias towards $h_1-J$ that happens for $\sigma_2=-1$. The surprising fact is that the temporal bias towards $t_+$ is exactly canceled by the smaller effect of $\sigma_2=+1$ on the magnetization of $\sigma_1$. This exact cancellation is special for models that have the echo-chamber symmetry, as we discuss below.

\subsection{The Blume-Capel model displays an echo chamber}\label{Sec:BlumeCapel}
We have seen a simple example of the lack of echo chambers in the Ising model and provided a heuristic argument to explain it. However, the exact cancellation between the fluctuations and bias in the Ising model is somewhat special and does not exist in other models. We, therefore, demonstrate the existence of an echo chamber with a second simple example. To this end, we consider a similar two spin system, but with spin-1 objects, instead of the spin-$\frac{1}{2}$ used in the Ising model. The spin-1 objects $\sigma$ can take the values $1$, $0$ or $-1$. We start with a single particle system. Its Hamiltonian is given by
\begin{eqnarray}
    \mathcal{H}(\sigma_1) = -h\sigma_1.
\end{eqnarray}
The partition function of this system is given by 
\begin{eqnarray}
    Z = e^{-\beta h} + e^{\beta h} + 1 = 2\cosh(\beta h) + 1.
\end{eqnarray}
The mean magnetization is given by
\begin{eqnarray}\label{Eq:BlumeCapelMag}
    \braket{\sigma_1} = \frac{\partial_h Z}{\beta Z} = \frac{\sinh(\beta h)}{\cosh(\beta h) + \frac{1}{2}},
\end{eqnarray}
which for $h>0$ is convex in $h$. Next, consider a system of two coupled spins of this type, namely a Hamiltonia
\begin{eqnarray}
    \mathcal{H}(\{\sigma\}) = -J\sigma_1\sigma_2 - h_1\sigma_1 - h_2\sigma_2.
\end{eqnarray}
The partition function is now
\begin{eqnarray}
    Z 
    &=&2\left(\cosh(\beta h_1) + \cosh(\beta h_2) + e^{\beta J}\cosh\left(\beta(h_1+h_2) \right) + e^{-\beta J}\cosh\left(\beta(h_1-h_2)\right) \right) + 1,
\end{eqnarray}
and the mean magnetization of say  $\sigma_1$ is given by 
\begin{eqnarray}\label{Eq:Mean_Sigma_1_coupled}
    \braket{\sigma_1} = \frac{2\left(\sinh(\beta h_1) + e^{\beta J}\sinh\left(\beta(h_1+h_2) \right) + e^{-\beta J}\sinh\left(\beta(h_1-h_2)\right) \right)}{2\left(\cosh(\beta h_1) + \cosh(\beta h_2) + e^{\beta J}\cosh\left(\beta(h_1+h_2) \right) + e^{-\beta J}\cosh\left(\beta(h_1-h_2)\right) \right) + 1}.
\end{eqnarray}
Using these we can easily check if there is a spin echo-chamber effect in the system. Namely, we can compare the mean magnetization of $\sigma_1$ at some field $h_1$ when the spin is coupled or uncoupled to the second spin with $h_2=0$. Plugging $h_2=0$ in Eq.(\ref{Eq:Mean_Sigma_1_coupled}) gives:
\begin{eqnarray}
    \braket{\sigma_1} 
    = \frac{\sinh(\beta h_1) }{\cosh\left(\beta h_1\right) + \frac{3}{2\left(1+ 2\cosh({\beta J})\right)}}
\end{eqnarray}
Since this is different from the single spin mean magnetization (Eq. \ref{Eq:BlumeCapelMag}), we see that this model indeed has an echo-chamber effect -- the mean magnetization is larger when the spin is connected to additional spin, and is only equal to it in the case $J=0$ where there is no interaction between the spins. It implies that while the different time fraction that $\sigma_2$ spends at each possible spin value decreases the echo chamber effect, it does not exactly compensate for the convexity of $\braket{\sigma(h)}$. Therefore, there is some degree of the echo chamber effect in the system.

\section{The Effective field}
The examples considered above give some insight into the echo-chamber effect. To further explore this effect in the different models, it is useful to introduce first the notion of an effective field which we present next.

Generally speaking, the mean magnetization of any specific spin at a fixed temperature $\beta$,  $\langle\sigma_k\rangle$, is a function of both the magnetic field on this spin, $h_k$, and the magnetic fields on all other spins, namely on $h_j$ for all $j\neq k$. Given the mean magnetization of spin $k$, we can invert Eq.(\ref{Eq:SingleSpinMean}) (or the corresponding relation between mean magnetization and external field for other models) and define the total effective field on this spin as:
\begin{eqnarray}
    h_{k}^{tot} = \beta^{-1}\tanh^{-1} \left(\braket{\sigma_k} \right).
\end{eqnarray}
$h_{k}^{tot}$ has the following physical interpretation: when $\sigma_k$ is decoupled from the network, $h_k^{tot}$ would make its mean magnetization identical to $\braket{\sigma_k}$. The mean magnetization $\braket{\sigma_k}$ is affected by both the actual field on the spin, $h_k$, and the interactions with the rest of the network. Thus, it is useful do define the ``network field'' which captures the latter, as 
\begin{eqnarray}\label{Eq:EffField_1}
    h_{k}^{network} = h_k^{tot}-h_k =  \beta^{-1}\tanh^{-1} \left(\braket{\sigma_k} \right) - h_k,
\end{eqnarray}
which can also be written as
\begin{eqnarray}\label{Eq:EffField_3}
    \langle\sigma_k\rangle = \tanh\left(\beta(h_k + h_k^{network})\right). 
\end{eqnarray}
An analogous network field can be defined for any spin model.  However, it is not a useful quantity in most cases (unless some echo-chamber symmetry discussed below exists in the system). This is because $h_k^{network}$ is, in general, a non-trivial function of $h_k$ itself. It implies that, for a general model, the external field on a spin influences the feedback that this spin receives from the network. Hence, the spin is subject to an ``echo-chamber''. 

In models where $h_k^{network}$ is independent of $h_k$, we call it the \emph{effective field} and denote it by $h^{eff}$. In these cases it can also be defined as 
\begin{equation}\label{Eq:EffField_2}
    h_{k}^{eff}=\beta^{-1}\tanh^{-1}(\braket{\sigma_k[h_k=0]}),
\end{equation}
namely as the total effective field when the external field on the specific spin is zero, and $[h_k=0]$ is short-hand notation for the external field configuration $(h_1,h_2,...,h_k=0,...h_N)$.  

As we next show, the Ising model displays this special property and, as a result, lacks a spin echo-chamber effect.

\subsection{Effective field in the Ising model}
In this subsection, we prove that one can define an effective field in the Ising model on an arbitrary network. In other words, we show that   $h_k^{eff}$, as defined in Eq.(\ref{Eq:EffField_1}),  is not a function of $h_k$. We do this by showing that Eq.(\ref{Eq:EffField_2}) holds, regardless of the value of $h_k$. To this end, we first define the following quantities:
\begin{equation}
    \mathcal{H}'_k(\vec\sigma) = \mathcal{H}(\vec\sigma)+\sigma_kh_k = -\sum_{\{i,j\}\in G}J_{ij}\sigma_i\sigma_j-\sum_{i\neq k} h_i \sigma_i,
\end{equation}
and 
\begin{eqnarray}
    A_+&=&\sum_{\{\sigma\neq \sigma_k\}}e^{\mathcal{-\beta H}'(\vec\sigma,\sigma_k=1)};\\
    A_-&=&\sum_{\{\sigma\neq \sigma_k\}}e^{\mathcal{-\beta H}'(\vec\sigma,\sigma_k=-1)},
\end{eqnarray}
where the summation in the definitions of $A_{\pm}$ is over all configurations of all the spins except $\sigma_k$, and $\mathcal{H}'(\vec\sigma,\sigma_k=\pm1)$ is calculated at the corresponding configuration but forcing $\sigma_k=\pm 1$, respectively. With this notation, we can write 
\begin{eqnarray} \label{Eq:Z_expressed_with_A}
    Z &=& e^{\beta h_k}A_+ + e^{-\beta h_k}A_-,\nonumber\\
    \partial_{h_k}Z &=& \beta\left(e^{\beta h_k}A_+ - e^{-\beta h_k}A_-\right) ,
\end{eqnarray}
and the mean magnetization can be written as
\begin{equation}\label{origin_mag}
    \braket{\sigma_k}=\frac{e^{\beta h_k} A_+-e^{-\beta h_k} A_-}{e^{\beta h_k} A_++e^{-\beta h_k} A_-}.
\end{equation}

Manipulating the above equation:
\begin{equation}
    \braket{\sigma_k}=\frac{e^{\beta h_k} A_+-e^{-\beta h_k}A_-}{e^{\beta h_k} A_++e^{-\beta h_k}A_-}=\frac{(e^{\beta h_k} -e^{-\beta h_k})(A_++A_-)+(e^{\beta h_k} +e^{-\beta h_k})(A_+-A_-)}{(e^{\beta h_k} +e^{-\beta h_k})(A_++A_-)+(e^{\beta h_k} -e^{-\beta h_k})(A_+-A_-)},
\end{equation}
where in the right hand side  $e^{-\beta h_k}A_+$ and $e^{\beta h_k}A_-$ were added and subtracted from both the numerator and denominator. Next, we multiply the numerator and the denominator by $\beta/2$, and  use the identity $A_++A_-=Z(h_k=0)$ as well as $\beta\left(A_+-A_-\right)=\partial_{h_k} Z(h_k=0)$, where $Z(h_k=0)$ is the partition function of the same system with $h_k$ set to be zero (See Eq. \ref{Eq:Z_expressed_with_A}). With these we get: 
\begin{equation}\label{effmag}
    \braket{\sigma_k}=\frac{\beta \sinh(\beta h_k)Z(h_k=0)+\cosh(\beta h_k)\partial_{h_k}Z(h_k=0)}{\beta \cosh(\beta h_k)Z(h_k=0)+\sinh(\beta h_k)\partial_{h_k}Z(h_k=0)}.
\end{equation}
Dividing numerator and denominator by $\cosh(h_k)\beta Z(h_k=0)$ and using Eq.(\ref{Eq:MeanSigma_k}), we get:
\begin{equation}
    \braket{\sigma_k}=\frac{\tanh(\beta h_k)+\braket{\sigma_k[h_k=0]}}{1+\tanh(\beta h_k)\braket{\sigma_k[h_k=0]}}.
\end{equation}
To continue, we use the following identity for hyperbolic tangent:
\begin{eqnarray}
    \tanh(a+b)=\frac{\tanh(a)+\tanh(b)}{1+\tanh(a)\tanh(b)},
\end{eqnarray}
to get:
\begin{equation}\label{Eq:UsingH_eff}
    \braket{\sigma_k}=\frac{\tanh(\beta h_k)+\braket{\sigma_k[h_k=0]}}{1+\tanh(\beta h_k)\braket{\sigma_k[h_k=0]}} = \tanh\left(\beta h_k + \tanh^{-1}\braket{\sigma_k[h_k=0]}\right).
\end{equation}
Lastly, we note that by definition (see Eq.\ref{Eq:EffField_3}), the above equation implies that 
\begin{eqnarray}
    h_k^{network} =\tanh^{-1}\braket{\sigma_k[h_k=0]}.
\end{eqnarray}
However, in this case, $h_k^{network}$ is calculated using the mean magnetization of the system when replacing the actual value of $h_k$ with $h_k=0$. This completes the proof that $h_k^{network}$ is $h_k$ independent. Therefore, in the Ising model $h_k^{network}$ can indeed be termed $h^{eff}_k$ as defined above.

We note that this result (and thus everything that follows), fails in the thermodynamic limit if the system has two distinct phases, where ergodicity breaking dictates averaging over a subspace of all micro-states, and one cannot evaluate the mean magnetization using Eq.(\ref{Eq:MeanSigma_k}). As we discuss in what follows, the property we just proved stems from a symmetry of the spin interactions. However, before showing this, we first explain why the existence of an effective field implies that the Ising model has no spin echo-chambers.

\subsection{The Ising model does not display spin echo-chambers}

As discussed above, a spin echo-chamber effect occurs when the coupling of a spin to a neutral network affects its response to an external magnetic field. To explain the effect, consider first the general case of a spin, $\sigma_1$ that can be coupled to a general network of spins $G$. When no external field is applied on any of the spins in the system, the average magnetization of $\sigma_1$ is zero (as is the case for any other spin in the system), namely $\braket{\sigma_1} = \braket{\sigma_k}=0$. We define a network to be \emph{neutral with respect to $\sigma_1$}, or \emph{$\sigma_1$-neutral}, in the following case: if no external field is applied on $\sigma_1$, its mean magnetization is zero regardless of it being connected or disconnected from the network. In other words, the connection to the network does not bias the mean value of $\sigma_1$. We note that by definition, a network is $\sigma_k$-neutral exactly when $h_k^{network}=0$ for $h_k=0$. If there is an effective field in this model, as is the case for the Ising model, then a network is $\sigma_k$-neutral when $h_k^{eff}=0$.  


Employing this definition and the effective field described above, it is now straightforward to show that there is no spin echo-chamber effect in the Ising model. 
Assume that a spin, which without loss of generality we denote by $\sigma_1$, is coupled to a $\sigma_1$-neutral network, and is subject to some external field $h_1$. By the effective field result, the mean magnetization of the spin is given by 
\begin{eqnarray}
    \langle\sigma_1\rangle = \tanh\left(\beta(h_1 + h_1^{eff})\right), 
\end{eqnarray}
but based on Eq.(\ref{Eq:EffField_2}) and the fact that the network is $\sigma_1$-neutral, it must be that $h_k^{eff}=0$ . Therefore,
\begin{eqnarray}
    \langle\sigma_1\rangle = \tanh\left(\beta(h_1 + h_1^{eff})\right) = \tanh(\beta h_1) = \braket{\sigma_1}_{uncoupled},
\end{eqnarray}
where $\braket{\sigma_1}_{uncoupled} = \tanh(\beta h_1)$ is the single spin equilibrium average of $\sigma_1$, namely its mean when it is not connected to any network. In other words, the above argument shows that the mean magnetization of the spin $\sigma_1$ is not affected by a connection to any $\sigma_1$-neutral network, and there is no spin echo-chamber effect in this model.

\subsection{Echo chambers in other models}
\label{Sec:OtherModel}
The argument used above to show that there are no echo-chambers in the Ising model,  can be applied to any model with the property that the network influence on a spin is independent of the local field. Namely, there is an effective field in the model. In models for which an effective field can be defined, the mean magnetization of $\sigma_1$ can be written as
\begin{eqnarray}
    \langle\sigma_1\rangle = f\left(\beta(h_1 + h_1^{eff})\right), 
\end{eqnarray}
where, as in the Ising model -- $h_1$ is the field on the spin $\sigma_1$, $h^{eff}_1$ is the influence of the network on the mean magnetization of $\sigma_1$ and it is independent on $h_1$, and $f$ is the function that relates the external field and the mean magnetization in an uncoupled spin for the specific model -- for example, in the Ising model $f(x) = \tanh(x)$. Since by its definition, the effective field is independent of $h_1$, then without loss of generality it can be calculated for the case $h_1=0$. But for $\sigma_1$-neutral network $h_1^{eff}=0$ in this case. Thus, we have in the above equation
\begin{eqnarray}
    \langle\sigma_1\rangle = f\left(\beta h_1 \right) = \braket{\sigma_1}_{uncoupled}, 
\end{eqnarray}
which implies that there is no spin echo-chamber effect in models for which an effective field can be consistently defined. This also implies that models with an echo-chamber effect would, in general, have no effective field as defined above. However, it is not always easy to understand if an effective field exists in the model or not -- in fact, we are not familiar with other models that have this property. Therefore, we next discuss a different approach towards investigating the existence of the echo-chamber effect.

To directly test the existence of an echo-chamber in a model, we compare two systems. In the first one there is a single, uncoupled spin. Its Hamiltonian is therefore given by
\begin{eqnarray}
    \mathcal{H}_1(\sigma_1) = -\bar h_1\cdot\bar \sigma_1.
\end{eqnarray}
 In the general case $\bar\sigma_1$ is a vector or a tensor, and $\bar {h}_1\cdot\bar{\sigma}_1$ is a scalar constructed from $\bar\sigma_1$ and the vector or tensor $\bar h_1$. The partition function, in this case is:
\begin{eqnarray}
    Z_1 = \sum_{\sigma_1} e^{\beta \bar\sigma_1\cdot \bar h}, 
\end{eqnarray}
where the sum signifies either discrete summation over all possible spin states (e.g.$\{-1,0,1\}$ for the case of spin $1$) or integration in the case of a continuous variable (e.g. angles for the $XY$ model).

We compare this scenario to a second scenario in which the spin $\bar\sigma_1$ is embedded in a network $G$ of spins with general coupling $J_{i,j}$ and fields $\bar h_i$. We denote the 
full Hamiltonian as 
\begin{eqnarray}
    \mathcal{H}_{N}(\vec\sigma) = -\sum_i \bar h\cdot\bar\sigma_i - \sum_{i,j} J_{i,j}\bar\sigma_i\cdot\bar\sigma_j,
\end{eqnarray}
 where again $\vec\sigma = (\bar\sigma_1,...\bar\sigma_N)$, and the partition function is 
 \begin{eqnarray}
    Z_N &=& \sum_{\bar\sigma_i} e^{\beta \bar\sigma_1\cdot\bar h_1 + \beta \sum_{i,j} J_{i,j}\bar\sigma_i\cdot\bar\sigma_j +\beta\sum_{i\neq 1} \bar h_i\cdot\bar\sigma_i} \nonumber\\
    &=&  \sum_{\bar\sigma_1}  e^{\beta \bar\sigma_1\cdot \bar h} \sum_{\sigma \neq \sigma_1} e^{\beta   \sum_{i,j} J_{i,j} \bar\sigma_i\cdot \bar\sigma_j +\beta\sum_{i\neq 1}\bar h_i\cdot\bar\sigma_i } \left(e^{\beta \bar\sigma_1  \cdot\sum_{\sigma_k \in G_1} J_{1,k} \bar\sigma_k}  \right).
\end{eqnarray}
We define \emph{strong echo-chamber  symmetric models} as models in which the term
\begin{eqnarray}\label{Eq:Z_eff_def}
   Z_{N-1}^{eff}(\bar\sigma_1) = \sum_{\{\bar\sigma_{k \neq 1}\}} e^{\beta   \sum_{i,j} J_{i,j} \bar\sigma_i\cdot\bar\sigma_j +\beta\sum_{i\neq 1} \bar h_i\cdot\bar\sigma_i} \left(e^{\beta \bar\sigma_1\cdot  \sum_{\bar\sigma_k \in G_1} J_{1,k} \bar\sigma_k}  \right) 
\end{eqnarray}
is independent of the specific value of $\sigma_1$ for every neutral network.  In such models, $Z_{N-1}^{eff}(\sigma_1)$ is not a function of $\sigma_1$ and it is therefore possible to factorise the partition function:
\begin{eqnarray}\label{Eq:Z_factorization}
    Z_N = \left(\sum_{\bar\sigma_1}  e^{\beta \bar\sigma_1\cdot\bar h_1}\right)\left( \sum_{\sigma \neq \sigma_1} e^{\beta   \sum_{i,j} J_{i,j} \bar\sigma_i\cdot\bar\sigma_j +\beta\sum_{i\neq 1} \bar h_i\cdot\bar\sigma_i} \left(e^{\beta \bar\sigma_1 \cdot \sum_{\sigma_k \in G_1} J_{1,k} \bar\sigma_k}  \right)\right) = Z_1\cdot Z_{N-1}^{eff}. 
\end{eqnarray}

This factorization of $Z_N$ implies that the magnetization of $\bar\sigma_1$ is independent of the existence of the rest of the network, and hence there is no echo-chamber effect in the model. To show this, we use, again, the relation between the mean magnetization and partition function in Eq.(\ref{Eq:MeanSigma_k}): 
\begin{eqnarray}
    \braket{\sigma_1} = \frac{\partial_{\bar h_1} Z}{\beta Z} = \frac{\partial_{\bar h_1} Z_1\cdot Z_{N-1}^{eff}}{\beta Z_1\cdot Z_{N-1}^{eff}} = \frac{\partial_{\bar h_1} Z_1 }{\beta Z_1 },
\end{eqnarray}
which is identical to the single spin magnetization, and therefore implies that there is no echo-chamber effect in the model.

As we have already seen by using the effective field method, the Ising model is a rare example of a model with a strong echo-chamber symmetry. However, there are several models with a \emph{weak echo-chamber symmetry}: in such models $Z_{N-1}^{eff}$ is $\bar\sigma_1$ independent for networks where $h_i=0$ for all $i\neq 1$. In such models, There are no echo-chambers for spins connected to networks that do not have additional local fields. Nevertheless, they may still display an echo-chamber effect in $\sigma_k$-neutral networks where the magnetic fields on different spins balance each other out. 
We next discuss each of these different cases.

\subsubsection{Models with strong Echo-chamber symmetry}
We first show that the Ising model has strong echo-chamber symmetry. We do this without employing effective field considerations, which turns out to be instructive when considering other models. To this end, we use the fact that the network is $\bar\sigma_1$-neutral. Therefore 
\begin{eqnarray}
    \braket{\bar\sigma_1}=\left.\frac{\partial_{\bar h_1}Z_N}{\beta Z_N}\right|_{h_1=0,h_2,h_3...h_N}=0,
\end{eqnarray}
which implies that 
\begin{eqnarray}
    \left.\beta^{-1}\frac{\partial Z_N}{\partial{\bar h_1}}\right|_{h_1=0,h_2,h_3...h_N}=0.
\end{eqnarray}
Substituting the expression for the partition function, we get that:
\begin{eqnarray}\label{Eq:StrongSymmetry}
    \sum_{\sigma_1}  \bar\sigma_1 \sum_{\sigma \neq \sigma_1} e^{\beta   \sum_{i,j} J_{i,j} \bar\sigma_i\cdot\bar\sigma_j +\beta\sum_{i\neq 1} \bar h_i\cdot\bar\sigma_i } \left(e^{\beta \bar\sigma_1 \cdot \sum_{\sigma_k \in G_1} J_{1,k} \bar\sigma_k}  \right)=\sum_{\sigma_1}  \sigma_1 Z_{N-1}^{eff}(\sigma_1) =0.
\end{eqnarray}
Carrying out the sum over the two possibilities $\sigma_1=\pm 1$, we obtain: 
\begin{eqnarray}
     Z_{N-1}^{eff}(\sigma_1=+1)=Z_{N-1}^{eff}(\sigma_1=-1),
\end{eqnarray}
and therefore a neutral network implies that $Z_{N-1}^{eff}$ is not a function of $\sigma_1$, hence there is a strong echo-chamber symmetry in this model.

We note, however, that we do not expect many other models to be in the class of the strong echo-chamber symmetry, as in general, Eq.(\ref{Eq:StrongSymmetry}) does not imply Eq.(\ref{Eq:Z_factorization}) holds. Therefore, neutral networks in other models do not generally imply the separability of the partition function.

\subsubsection{Models with  weak Echo-chamber symmetry}
Whereas a $\sigma_k$-neutral network is not enough to eliminate the echo-chamber effect in most models, a special case of interest is when the network is $\sigma_k$-neutral because there are no magnetic fields in the system on any spin except $\sigma_k$. The class of models that have the echo-chamber symmetry in this special case but not for general $\sigma_k$-neutral networks, which is the class of weak echo-chamber symmetric models, includes models with the following spin-symmetry: for any change $\sigma_1\to\hat\sigma_1$, there exist a one-to-one transformation $\hat\sigma_k = F_{\sigma_1\to\hat\sigma_1}[\sigma_k]$ such that $\sigma_i\sigma_j = \hat\sigma_i\hat\sigma_j$ for any $i$ and $j$. Mathematically, this implies a symmetry group $F$ of invertible transformations on the possible values of $\sigma$, such that the spin multiplication in the Hamiltonian is invariant under the action of elements of $F$. In this case, $Z_{N-1}^{eff}(\sigma_1)$ is $\sigma_1$ independent regardless of the specific values of $J_{ij}$. 

To show that models with a symmetry transformation $F$ have the weak echo-chamber symmetry,  we note that a transformation $\sigma_1\to\hat\sigma_1$ generally changes all the elements in the sum over $\{\sigma_{k\neq 1}\}$ in the definition of $Z_{N-1}^{eff}$ (Eq. \ref{Eq:Z_eff_def}), but each of them is equal to another element in the sum, given by the realization corresponding to the transformation in which $\sigma_k \to F_{\sigma_1\to\hat\sigma_1}[\sigma_k]$. Therefore, the contribution of each configuration $\{\sigma_{k\neq 1}\}$ with $\sigma_1$ is equal to the contribution of the configuration $\{\hat\sigma_{k\neq 1}\}$ with $\hat\sigma_1$. This is a result of the multiplication between spins being invariant under any group action, so replacing $\sigma_i\sigma_j$ with $F_{\alpha}\sigma_iF_{\alpha}\sigma_j$  in $Z_{N-1}^{eff}$ does not change its value for any $F_{\alpha}$ in the group. Therefore, the total sum is independent of the value of $\sigma_1$. In other words, changing the value of $\sigma_1$ reshuffles the order of the elements in the sum, but does not change the summation value. 
We note, however, that even models that have no such a transformation might nevertheless have the echo-chamber symmetry for some specific realizations of $J_{ij}$.   

An example for a model with a weak echo-chamber symmetry is the $XY$ model, where $\sigma_i = (\cos\theta_i,\sin \theta_i)$. In this case, any change of $\sigma_1\to\hat\sigma_1$ corresponds to some change in the angle $\theta_1\to\hat \theta_1 = \theta_1+\Delta\theta$. We can define the transformation $$\hat \theta_k = F_{\theta_1\to\theta_1+\Delta\theta}[\theta_k] = F_{\Delta\theta}\theta_k= \theta_k+\Delta\theta,$$ where $F_{\Delta\theta}$ is a rigid rotation of a spin by the corresponding angle $\Delta\theta$. In this case the symmetry group is $SO(2)$, and it preserves the multiplication in the Hamiltonian. 
In the absence of external fields, a realization $\{\theta_{k\neq 1}\}$ contributes with $\theta_1$ exactly as the realization $\{\hat\theta_{k\neq 1}\}$ with $\hat\theta_1$, and therefore the contribution of $Z_{N-1}^{eff}$ is independent of the specific angle of $\sigma_1$. In this case, a change in $\sigma_1$ can be viewed as a change in the reference frame, which does not change $Z_{N-1}^{eff}$. However, this is not the case in the presence of external fields $\vec h$, and therefore the model has only the weak but not the strong echo-chamber symmetry. We note that any model with a similar symmetry group that respects the two spins interaction (e.g., the Heisenberg model and the vector Potts model \cite{wu1982potts}) has a weak echo-chamber symmetry that protects the model from echo-chambers in the absence of external fields. However, we also note that models with weak echo chamber symmetry generally do not have an effective field, as this would imply that they have the strong rather than the weak echo-chamber symmetry.

\subsubsection{Models without Echo-chamber symmetry}
Last, we note that many models of interest have neither the strong nor the weak echo-chamber symmetry. Correspondingly, these models have an echo-chamber effect for most realizations of neutral networks. An example for such a model is the spin-1 Blume-Capel model \cite{blume1966theory,capel1966possibility}, where an echo chamber effect was explicitly demonstrated in Sec. \ref{Sec:BlumeCapel} .

\section{Precise calculation of effective field and magnetization on tree networks}
In the previous section, we have shown that, for models with a strong echo-chamber symmetry, the mean magnetization of each spin can be calculated using an effective field approach, where the effective field is not a function of the field on the spin itself. In this section, we show that in the case of the Ising model on tree networks, the effective field can be calculated using a relatively simple, closed formula. We then introduce an efficient algorithm that uses this formula to calculate the mean magnetization on each spin and the entire system.

\subsection{Effective Field Formula}

\subsubsection{Leaf Spins:}
To explain how the effective field developed above can be used to construct an efficient algorithm to calculate the mean magnetization on a tree, consider first a ``leaf'' edge on a tree, namely a spin connected to only one other spin. Without loss of generality, let us denote the leaf spin as $\sigma_1$, and the spin connected to is as $\sigma_2$. Using the effective field approach, we know that 
\begin{eqnarray}
    \braket{\sigma_1}=\tanh\left(\beta(h_1+h^{eff}_1)\right),
\end{eqnarray}
and therefore our aim is to calculate $h^{eff}_1$, which can be done by calculating $\braket{\sigma_1}$ with $h_1=0$. The latter is given by:
\begin{equation}
    \braket{\sigma_1}[h_1=0]=\frac{\sum_{\{\sigma\}}\sigma_1 e^{\beta(\sum_i h_i\sigma_i+\sum_{\{i,j\}\in G}J_{ij}\sigma_i\sigma_j )}}{\sum_{\{\sigma\}}e^{\beta(\sum_i h_i\sigma_i+\sum_{\{i,j\}\in G}J_{ij}\sigma_i\sigma_j )}}.
\end{equation}
Next, we perform the summation over $\sigma_1 = \pm 1$, and get:
\begin{equation}
    \braket{\sigma_1}[h_1=0]=\frac{\sum_{\{\sigma_k\neq \sigma_1\}} e^{\beta(\sum_{i\neq 1} h_i\sigma_i+\sum_{\{i\neq 1,j\neq 1\}\in G}J_{ij}\sigma_i\sigma_j )}\left(e^{\beta J_{1,2}\sigma_2} - e^{-\beta J_{1,2}\sigma_2}\right)}{\sum_{\{\sigma_k\neq \sigma_1\}} e^{\beta(\sum_{i\neq 1} h_i\sigma_i+\sum_{\{i\neq 1,j\neq 1\}\in G}J_{ij}\sigma_i\sigma_j )}\left(e^{\beta J_{1,2}\sigma_2} + e^{-\beta J_{1,2}\sigma_2}\right)}.
\end{equation}
We note that the term $\left(e^{\beta J_{1,2}\sigma_2} + e^{-\beta J_{1,2}\sigma_2}\right)$ in the denominator is independent on the specific value of $\sigma_2$, and is equal to $2\cosh(\beta J_{1,2})$. Similarly, the term $\left(e^{\beta J_{1,2}\sigma_2} - e^{-\beta J_{1,2}\sigma_2}\right)$ is equal to  $2\sigma_2\sinh(\beta J_{1,2})$ as $\sigma_2$ can only be $\pm 1$. We can therefore write: 
\begin{equation}
    \braket{\sigma_1}[h_1=0]=\tanh(\beta J_{1,2})\frac{\sum_{\{\sigma_k\neq \sigma_1\}} \sigma_2 e^{\beta(\sum_{i\neq 1} h_i\sigma_i+\sum_{\{i\neq 1,j\neq 1\}\in G}J_{ij}\sigma_i\sigma_j )}}{\sum_{\{\sigma_k\neq \sigma_1\}} e^{\beta(\sum_{i\neq 1} h_i\sigma_i+\sum_{\{i\neq 1,j\neq 1\}\in G}J_{ij}\sigma_i\sigma_j )}} = \tanh(\beta J_{1,2})\braket{\tilde \sigma_2}
\end{equation}
where $\braket{\tilde \sigma_2}$ is, by the expression in the middle, the mean magnetization of $\sigma_2$ in a system where the spin $\sigma_1$ is deleted, namely the leaf is trimmed from the graph. In terms of the effective field, we can write:
\begin{eqnarray}\label{Eq:h_1_eff_formula}
    h_1^{eff} = \beta^{-1}\tanh^{-1}\Big(\tanh(\beta J_{1,2})\braket{\tilde \sigma_2}\Big)
\end{eqnarray}

This structure, in which $h_1^{eff}$ is calculated from a trimmed version of the same tree, naturally calls for a recursion process. However, caution must be made when using the above procedure, since (say in the above example) $\sigma_2$ may be connected to more than two spins, so trimming $\sigma_1$ does not make $\sigma_2$ a leaf spin. As we next show, this result can be generalized to non-leaf spins. 

\subsubsection{Non-Leaf Spins:}
Consider a non-leaf spin; namely, it is connected to several spins. Let us denote the specific spin by $\sigma_k$, and its neighboring spins by $\sigma_m$...$\sigma_n$. Since the spin interaction network is a tree graph, deleting the node associated with $\sigma_k$ would make each of the spins connected with it $\sigma_m$...$\sigma_n$, part of a disconnected tree. We denote these trees by the $G_m$...$G_n$ correspondingly, and with this notation, we can write the Hamiltonian as:
\begin{eqnarray}
    \mathcal{H}(\{\sigma\}) =   \left(\sum_{i\in\{m...n\}}\mathcal{H}_i\left(\{\sigma\}\right) - J_{ki}\sigma_k\sigma_i\right) - h_k\sigma_k, 
\end{eqnarray}
where 
\begin{eqnarray}
    \mathcal{H}_i(\{\sigma\}) = -\sum_{\{l,j\}\in G_i}J_{lj}\sigma_l\sigma_j - \sum_{j\in G_i}h_j\sigma_j.
\end{eqnarray}
In other words, $\mathcal{H}_i(\{\sigma\})$ is the contribution to the Hamiltonian from the sub-tree $G_i$.

Using the notation above, the mean magnetization of the non-leaf spin can be written as:
\begin{equation}
    \braket{\sigma_{k}}[h_{k}=0]=\tanh\left(\sum_{h\in \{m...n\}} \tanh^{-1}(\tanh(\beta J_{hk})\braket{\tilde\sigma_h})\right),
\end{equation}
which in terms of the effective field can be written as:
\begin{eqnarray}\label{Eq:h_eff_complete}
    h_k^{eff} = \beta^{-1}\sum_{h\in\{m...n\}}\tanh^{-1}\Big(\tanh(\beta J_{hk})\braket{\tilde \sigma_h}\Big).
\end{eqnarray}
In the above equation, $\braket{\tilde \sigma_h}$ is the mean of $\sigma_h$ calculated using $\mathcal{H}_h$, namely as if we cut the connection between the spins $k$ and $h$. The proof of this result involves cumbersome, though straightforward algebra, and is therefore left to the appendix (See appendix \ref{Appendix_NonLeafProof}).

As in the case of the leaf spin, here too the mean magnetization and effective field on the spin $k$ can be calculated from the mean magnetization of the sub-trees connected to the spin $k$. This enables a recursive and efficient calculation of the mean magnetization of each spin, as described in the next section.

\subsection{Algorithm for precise magnetization calculation}\label{Sec:Algo}

The recursion formula for the effective field, as formulated above, allows fast and precise calculation of the mean magnetization of any spin and over the tree network as a whole. For clarity of presentation, the algorithm described below is sub-optimal in terms of running time and memory allocation but is simpler to implement. In this algorithm, we first reshape the tree using the Breadth-first search (BFS) algorithm \cite{silvela2001breadth}, such that the calculations only involve leaf nodes, and never the more complicated case, whereby trimming the node the tree becomes a set of disconnected trees.

To calculate the mean magnetization of a given spin, the algorithm first reshapes the tree such that this spin is the tree root, using the BSF algorithm, which for completeness is provided as a pseudo-code in Appendix \ref{Sec:AppAlgo} (Algorithm 3). The effective field is then calculated recursively (Algorithm 2 in Appendix \ref{Sec:AppAlgo}), starting from the leaves of the resulting tree, removing them after taking into account their effective field and its influence on their neighbors using Eq.(\ref{Eq:h_eff_complete}),  and then repeating the same process with the remaining tree until the root is the only node in the graph. By scanning the mean magnetization of all the spins (Algorithm 1 in Appendix \ref{Sec:AppAlgo}), the mean magnetization over the network can be calculated.

\begin{figure}[h]
  \centering
    \includegraphics[width=0.8 \textwidth]{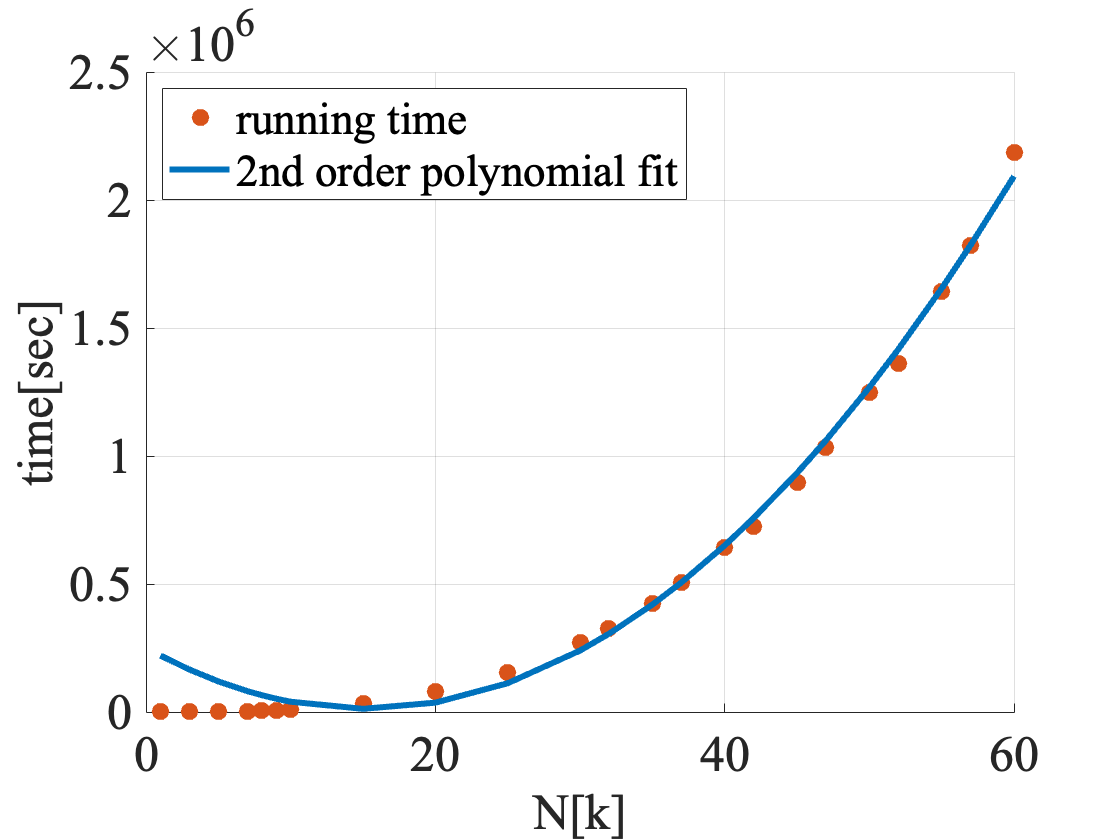}
\caption{Run time of the algorithm as a function of the number of spins in the tree (in thousands). As expected, an $\mathcal{O}(N^2)$ dependence is observed.}
\label{fig:comp_t}
\end{figure}

The time complexity of this procedure can be compared to the complexity of the direct calculation of the total magnetization by averaging over all microstates, namely all configurations of the system. The time it takes to calculate the mean magnetization in the latter case scales as the number of microstates of the networks, therefore it grows at least exponentially with the number of spins $N$, namely it is $\mathcal{O}(e^N)$. On the other hand, the algorithm presented above includes, for each node, a BFS step required to calculate the depth and parents structure given a specific root and then backtracking over the resulting structure to calculate the effective field. Both of these  scale with the number of nodes in the system as $\mathcal{O}(N)$ \cite{silvela2001breadth}. Since this process is repeated for each node the resulting time complexity is $O(N^2)$  (shown in figure \ref{fig:comp_t}), which is significantly faster than direct averaging over the microstates. The algorithm may be made more efficient if one exploits the fact that the trees with nearby roots are very similar in structure. We leave such improvements to later work. 

To summarize, the lack of echo-chamber effect in the Ising model translates into a significant decrease in the complexity for calculating the mean magnetization.

\section{Employing the effective field algorithm }

In this section, we demonstrate the strength of the method presented above on two examples that incorporate large tree networks with hundreds of nodes.

\subsection{Random Field Ising Model on the Bethe Lattice}
\begin{figure}[h]
  \centering
    \includegraphics[width=0.49 \textwidth]{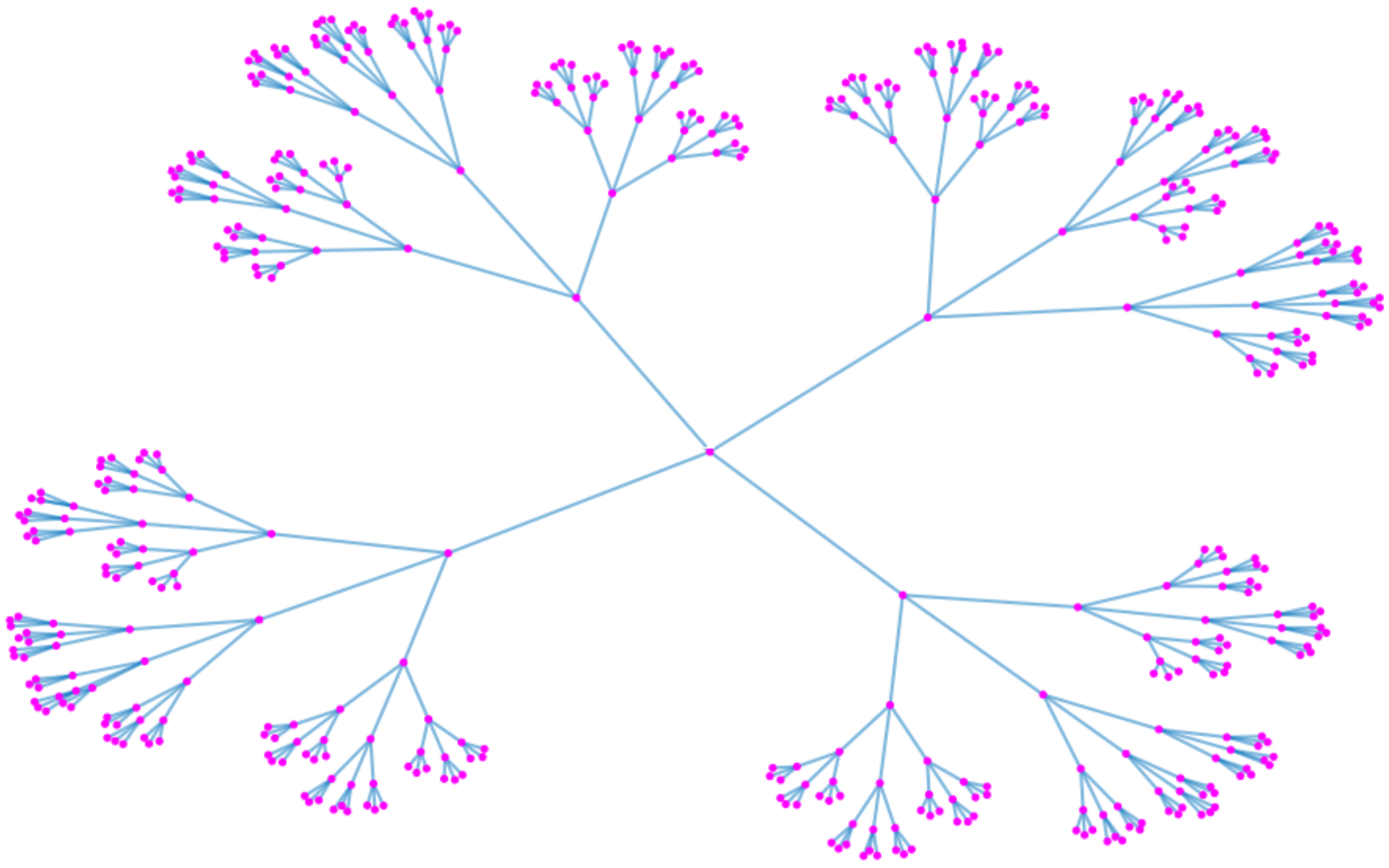}
    \includegraphics[width=0.49 \textwidth]{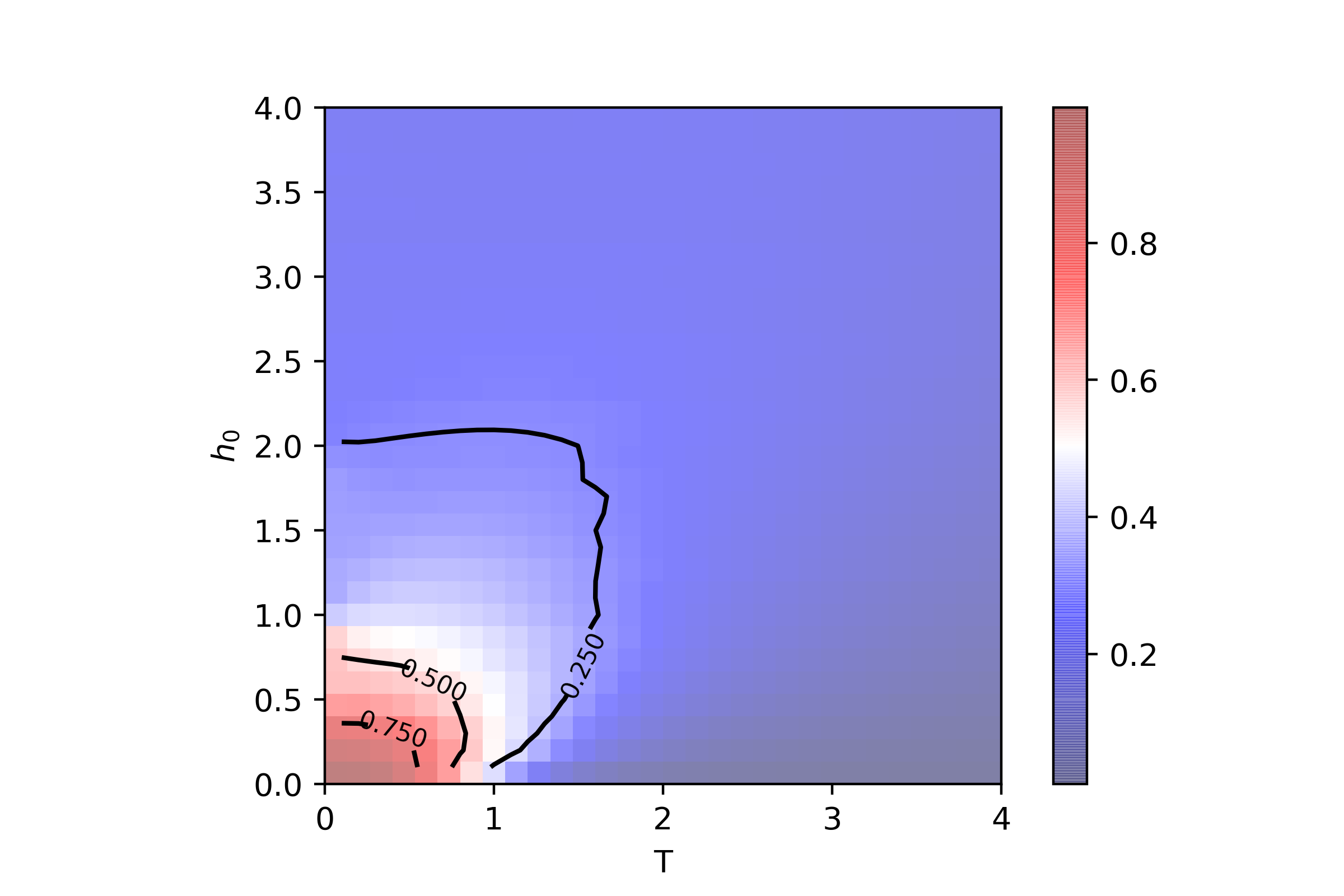}
\caption{RFIM on the Bethe lattice. Left panel: the Bethe lattice with coordination number 4 and 5 shells on which the mean magnetization was calculated. Right panel: the mean magnetization on this lattice, with random fields sampled from a Gaussian distribution with variance $h_0$ and temperature $T$.}
\label{fig:3dPT}
\end{figure}

Let us demonstrate the algorithm developed above on the famous problem of the Random Field Ising Model (RFIM) on the Bethe lattice \cite{bruinsma1984random,nowotny2001phase,bleher1998phase}. The RFIM is described by the following Hamiltonian,
\begin{eqnarray}
    \mathcal{H} = -J\sum_{\{i,j\}\in G}\sigma_i \sigma_j - \sum_i h_i \sigma_i .
\end{eqnarray}
Specifically, we are interested in the case where $G$ is the Bethe lattice, a tree-graph where each spin is connected to the same number of other spins. The number of neighbors each spin has in the tree is called \emph{the coordination number} of the graph. The left panel of Fig.(\ref{fig:3dPT}) depicts the Bethe lattice with coordination number 4. In the thermodynamic limit, this model exhibits a phase transition from a ferromagnetic phase for small random fields to a paramagnetic phase for large fields \cite{bruinsma1984random,nowotny2001phase}. It is, therefore, a good example where averaging the magnetization over different realizations of the quenched disorder can be useful. To this end, we performed the following calculation on a Bethe lattice with 485 spins, coordination number 4 (each spin in the bulk has four neighbors), and five shells, namely from the center of the graph there are five steps until the boundary is reached (see Fig.\ref{fig:3dPT}). For each value of the temperature $T$ and standard deviation $h_0$ of the random fields $h_i$, we chose 100 different realizations of $\vec h$, and for each of these  realizations the mean magnetization in the lattice was calculated. The right panel in Fig.(\ref{fig:3dPT}) shows the quenched mean magnetization. We note that in order not to average positive and negative magnetization in the ordered phase, we average over $m=|N^{-1}\sum_i\sigma_i|$, but in a finite system, the absolute value implies that $m$ is not zero, but rather scales as $N^{-0.5}$ even in the disordered phase. 

In this example, we used the algorithm developed above to exactly calculate the magnetization in each realization, as a ``tour de force'' application on a system where much is known on its phase diagram \cite{nowotny2001phase}. However, the algorithm is quite general and can be used on many other applications, as we show below.

\subsection{Influence maximization} 
In this section, we use the algorithm introduced earlier to solve a problem with a similar flavor to the echo-chamber effect, which is the main motivation in this manuscript: influence maximization for large social network \cite{liu2010influence,lynn2016maximizing,lynn2017statistical}. In this problem, the goal is to find a small seed in a social network that could maximize the spread of influence. We assume a network of agents with two competing opinion states, such that ``+1'' represents the preferred opinion, and ``-1'' the opposite opinion. The agents have social interaction that tends to ``align" their opinions; namely, an agent tends to change its opinion towards the mean opinion of the agents it is socially connected with. Two additional influences on each agent are: (i) some internal tendency to switch opinion, which we assume to be uniform among all agents. This mechanism is commonly modeled as random fluctuations induced by the physical temperature of the system; (ii) External influence on each agent that can come from advertisements or any other source. This influence is commonly modeled as an external magnetic field on each network node.  

Such problems are commonly analyzed using the ferromagnetic Ising model described above, where the mean opinion in the network is given by the mean magnetization at thermal equilibrium \cite{brede2019transmission,lynn2016maximizing,lynn2017statistical,lynn2018maximizing,moreno2019shielding,romero2020continuous}. The main question is then as follows: given a limited budget of external influence (say a limit budget for advertisement), modeled as a limited amount of external field denoted by $$H=\sum_i h_i,$$ where $h_i$ is the local external influence on the site $i$ and $H$ is the total external influence, how should one partition the external influence between the different spins as to maximize mean network magnetization. In other words, find the set $\{h_i\}$ that maximizes $\braket{m}=\braket{\sum_i\sigma_i}$ under the constraints $\sum_ih_i=H$ and $h_i>0$. Mathematically, this can be phrased as:
\begin{equation}\label{mag_der}
    h_i = \arg  \underset{h_i>0}{\text{max }}  \text{ } \langle m[h_i,J,H,\beta]\rangle
\end{equation}
\begin{equation*}
    \text{ s.t. } \sum_{i=0}^{N}h_i=H
\end{equation*}

\begin{figure}[h] 
  \centering
  \begin{minipage}[b]{0.45\textwidth}
    \includegraphics[width=\textwidth]{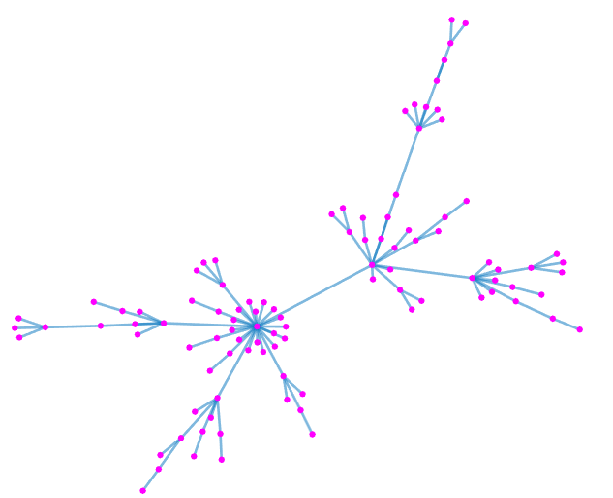}
  \end{minipage}
    \hfill
  \begin{minipage}[b]{0.5\textwidth}
    \includegraphics[width=\textwidth]{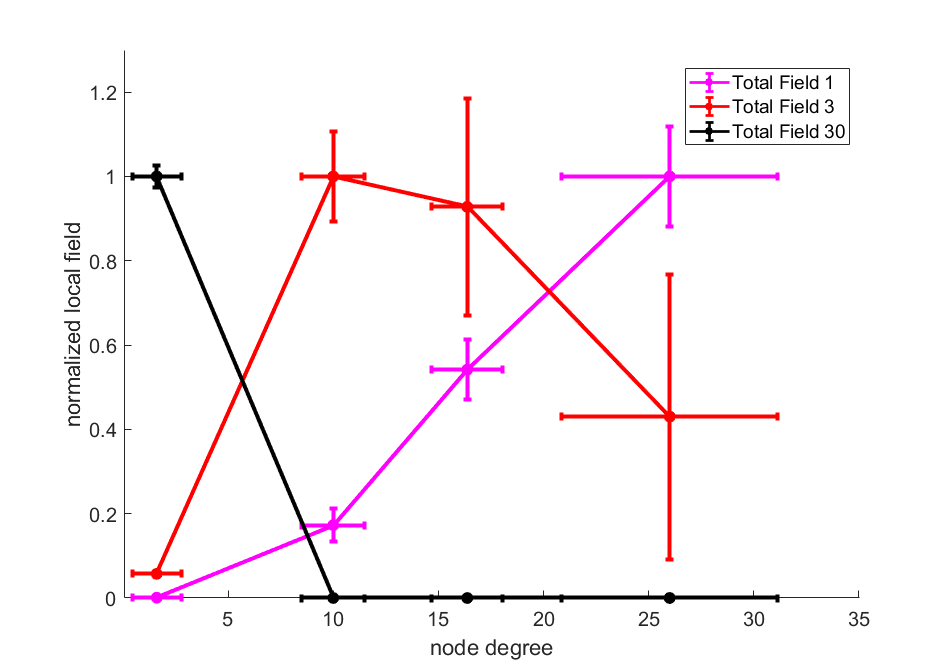}
  \end{minipage}
 \caption{ Influence maximization on a social network. Left panel: A sample of a scale-free network with $N=100$ nodes. The graph was generated using the Barabashi-Albert model. Right panel: Optimal local field for each node in the network, averaged over 12 different graphs (with $N =100$ nodes each, altogether 1200 nodes). The different colors stand for different value of the total field H (see legend). The results are binned by node degree (the x-axis), allowing us to compare data from networks with different degree distributions. The horizontal error bars are the standard deviation of the degrees of the nodes around each specific bin. We find that the optimal field shift from focusing on the hubs at the weak total field to the leaves at the strong total field. The vertical error bars represent the standard error of the mean across all nodes in a bin and show a significant deviation in the local field between nodes with similar (as well as equal) degrees. Since we consider a power-law graph, the number of nodes with a high degree is much smaller than the number of nodes with a low degree, which means that for H=1 (magenta) the field on the hubs is macroscopic, while for H=30 (black) where the field focus on the leaves, it is divided between many nodes, thus for each leaf, it is quite small.}\label{fig:BAopt}
\end{figure}

This problem was already addressed in the literature, using mean-field approximation or numerical simulations \cite{brede2019transmission,lynn2016maximizing,lynn2017statistical,lynn2018maximizing}. In the limit of weak field (small $H$), it was shown that the optimal field is strongest at the hubs. Conversely, the optimal field is stronger on the leaves in the strong field limit (large $H$). Using the above method, we calculated the exact magnetization on a large network for different values of the magnetic field on each node. 
Using the recursive algorithm introduced in Sec.\ref{Sec:Algo}, we can efficiently calculate the exact value of the total magnetization for each set of local fields $h_i$. Therefore, we could use the Sequential Linear/Quadratic Programming (SLQP) method for optimization \cite{leyffer2010nonlinear}, with which we found the set  $\{h_i^{opt}\}$ that maximizes magnetization for a given total field H. Here we choose to work with a Barabasi-Albert random graph, where the degree (the number of neighbors each spin has) is distributed as a power-law \cite{barabasi2014network}, to decrease the probability for loops. The specific graphs we chose (see example in figure \ref{fig:BAopt}) do not contain any loops, so the solution provided by the recursive method is the exact solution.

In general, one may still use this method to approximate non-tree graphs. In such cases, when the algorithm encounters a node that closes a loop, it considers it as connected separately to the first neighbor and then separately to the second one. The second-order connection between the neighbors are ignored, which provide a small correction that can be negligible in the case of very few loops. 

Earlier work \cite{brede2019transmission,lynn2016maximizing,lynn2017statistical,lynn2018maximizing,moreno2019shielding,romero2020continuous} has shown that the optimal field should be stronger for the high-connectivity nodes where the total field H is small and stronger on the lower-degree nodes where H is large. In figure \ref{fig:BAopt} we plot the normalized optimal field for each node vs. the node's degree for three different values of the total field H. The presented results depict averages over $R=12$ realizations of  Barabasi-Albert graphs with $N=100$ spins and overall nodes with similar degrees (error bars represent the standard error of the mean). As expected, when the total field is weak (H=1) the optimal field is concentrated mostly on the hubs, and when the total field is stronger (H=30), it is concentrated mostly on the leaves. For an intermediate fields strength (H=3), we find that the optimal influence is obtained by concentrating the external field on nodes of intermediate degree (rather than dividing it it between hubs and leaves).  Note that these results are general, and are not limited to the BA network \cite{brede2019transmission,lynn2016maximizing,lynn2017statistical,lynn2018maximizing,moreno2019shielding,romero2020continuous}. The assumption in previous work was that the qualitative behavior of the optimal field depends mostly on a node's degree and less so on the structure of the network. Our results show that, even though the average behavior matches with the above assumption, there is significant variation between the optimized local fields on different nodes with the same degree. Our method does not require symmetry assumption and allows for more accurate influence maximization schemes rather than general behaviors. 

We note in passing that the dual problem, namely influence minimization, has a somewhat simpler solution: the convexity of the mean magnetization of a spin as a function of the field on it implies that focusing all the field on the spin that has the least effect on the network minimizes the influence of the field on the network.

\section{Conclusions}
In this manuscript, we have shown that there is no echo-chamber effect in the Ising model. For other spin models, we identified symmetry that forbids the existence of echo-chambers. This symmetry can be strong -- as in the Ising model, and then it prevents any echo-chambers, or it can be weak -- in this case echo-chambers may exist, but only when there are additional fields in the system. A consequence of the lack of echo-chambers in the Ising model is that the mean magnetization of such networks can be efficiently solved in tree graphs, based on effective field calculations. We used this technique to construct an algorithm for finding the mean magnetization in tree graphs. We demonstrated it in two applications: the RFIM on the Bethe lattice and influence maximization. 

When an external field is applied on a spin, it biases the spin itself and all the spins connected to it towards the states that minimize the energy at this field. Nevertheless, this bias does not generate echo-chambers in the Ising model. The lack of echo-chamber effect and the existence of an effective field, in this case, are a consequence of the exact cancellation of two different effects: (i) Even in the presence of the external bias, the states of neighboring spins fluctuate, such that they spend more time in the low energy state; (ii) The low energy states of the neighboring spins generate a smaller bias on the spin, in comparison to the high energy states. The fact that these two factors exactly cancel each other is not guaranteed in all models. In fact, the Ising model is a unique case, such that this cancellation occurs even in the presence of additional fields. Other models, such as the XY, Heisenberg, and the vector Potts models,  display a weak echo-chamber symmetry, where such cancellations occur only in the absence of external fields in the network.     

The existence of an effective field in the Ising model can be interpreted as a stronger version of a ``no echo-chambers'' result, since it implies that an effective field can describe the impact of the network on the spin, $h^{eff}_1$, which is independent of the field on the specific spin, $h_1$. This is true even if the network is not $\sigma_1$-neutral, and therefore can be viewed as a generalization of the no echo-chambers result: the mean magnetization of the spin does not feedback through the network to change its magnetization, even if the network is not neutral for that spin.  

The expression for the effective field turned out to have a simple recursion structure in tree networks. Although we have shown that the effective field exists even in non-tree graphs (namely graphs that have loops), the general expression of the effective field for such graphs is not yet known. Extending our result to general graphs is expected to be quite difficult, as the problem of finding the mean magnetization of every spin in a general network is equivalent, in the low-temperature limit, to finding the ground state of the network -- a problem which is known to be NP-complete \cite{barahona1982computational}. Similarly, the separability of the partition function in Eq.(\ref{Eq:Z_factorization}) only holds for $\sigma_1$-neutral networks; therefore, generally, it cannot be applied iteratively since $Z^{eff}_{N-1}$ does not correspond to a neutral network.   

Applications of our results for physical systems, especially magnets, are of interest: do physical magnets display an echo-chamber effect? This question is interesting in both macroscopic systems, where ergodicity breaking implies that our results do not hold, as well as in microscopic systems (e.g., trapped ions \cite{kotler2014measurement}) where applying a magnetic field on each spin separately is often used, and the existence or non-existence of echo-chamber effect is therefore of interest. 

A natural question that we did not address concerns the cases where a spin echo-chamber effect exists, as in the spin-1 Blum-Capel model: does the existence of the effect imply that the mean magnetization of a spin coupled to a $\sigma_k$-natural network is always larger than the mean magnetization of a single spin (as in the example in section \ref{Sec:BlumeCapel}), or is it possible that coupling with a $\sigma_k$-natural network reduces the mean magnetization of $\sigma_k$? Additional questions of similar interest are the existence of the effect in non-equilibrium systems (e.g., a network with different temperatures for different spins) which is natural for influence maximization problem, or in the thermodynamic limit where phase transitions and the corresponding ergodicity breaking might change our results even in models that do have the echo-chamber symmetry defined above. It would also be of interest to understand the role of echo-chamber effect on opinion dynamic models, for example, by comparing models with and without the echo-chamber symmetry.    

We note that the algorithm we constructed is not optimal in terms of run-time, memory allocation, or any other standard performance measure. For example, it might be improved by optimizing its traverse over the tree when calculating the effective field.

Lastly, we note that the Ising model is quite special in that it lacks echo-chambers regardless of the network: most other models do contain echo-chambers, even in neutral networks. This special feature must be considered when using Ising spins in modeling opinion dynamics, especially when echo-chamber effects can play an essential role in the results.

\section*{Acknowledgements}
We would like to thank David Mukamel for useful discussions. We thank Anton Charkin-Gorbulin for help with the BFS algorithm. O.R. is the incumbent of the Shlomo and Michla Tomarin career development chair, and is supported by the Abramson Family Center for Young Scientists, the Minerva and by the Israel Science Foundation, Grant No. 950/19. O.F. is the incumbent of the  of the Henry J
Leir Professorial chair and is supported by the Israel Science Foundation, Grant No. 1727/20, the Minerva Foundation, and the European Research Council (ERC) under the European Unions Horizon 2020 research and innovation program (Grant agreement No. 770964).


\begin{thebibliography}{10}

\bibitem{jamieson2008echo}
Kathleen~Hall Jamieson and Joseph~N Cappella.
\newblock {\em Echo chamber: Rush Limbaugh and the conservative media
  establishment}.
\newblock Oxford University Press, 2008.

\bibitem{currin2021depolarization}
Christopher Currin, Sebastian~Vallejo Vera, and Ali Khaledi-Nasab.
\newblock Depolarization of echo chambers by random dynamical nudge.
\newblock {\em arXiv preprint arXiv:2101.04079}, 2021.

\bibitem{cinelli2021echo}
Matteo Cinelli, Gianmarco De~Francisci Morales, Alessandro Galeazzi, Walter
  Quattrociocchi, and Michele Starnini.
\newblock The echo chamber effect on social media.
\newblock {\em Proceedings of the National Academy of Sciences}, 118(9), 2021.

\bibitem{dubois2018echo}
Elizabeth Dubois and Grant Blank.
\newblock The echo chamber is overstated: the moderating effect of political
  interest and diverse media.
\newblock {\em Information, communication \& society}, 21(5):729--745, 2018.

\bibitem{bruns2017australian}
Axel Bruns, Brenda Moon, Felix M{\"u}nch, and Troy Sadkowsky.
\newblock The australian twittersphere in 2016: Mapping the follower/followee
  network.
\newblock {\em Social Media+ Society}, 3(4):2056305117748162, 2017.

\bibitem{del2018echo}
Marc~Esteve Del~Valle and Rosa~Borge Bravo.
\newblock Echo chambers in parliamentary twitter networks: The catalan case.
\newblock {\em International journal of communication}, 12:21, 2018.

\bibitem{Huszare2025334119}
Ferenc Husz{\'a}r, Sofia~Ira Ktena, Conor O{\textquoteright}Brien, Luca Belli,
  Andrew Schlaikjer, and Moritz Hardt.
\newblock Algorithmic amplification of politics on twitter.
\newblock {\em Proceedings of the National Academy of Sciences}, 119(1), 2022.

\bibitem{pathria2011statistical}
Raj~Kumar Pathria and Paul~D Beale.
\newblock Statistical mechanics, 2011.

\bibitem{mobilia2007role}
Mauro Mobilia, Anna Petersen, and Sidney Redner.
\newblock On the role of zealotry in the voter model.
\newblock {\em Journal of Statistical Mechanics: Theory and Experiment},
  2007(08):P08029, 2007.

\bibitem{hartnett2016heterogeneous}
Andrew~T Hartnett, Emmanuel Schertzer, Simon~A Levin, and Iain~D Couzin.
\newblock Heterogeneous preference and local nonlinearity in consensus decision
  making.
\newblock {\em Physical review letters}, 116(3):038701, 2016.

\bibitem{liu2010influence}
Shihuan Liu, Lei Ying, and Srinivas Shakkottai.
\newblock Influence maximization in social networks: An ising-model-based
  approach.
\newblock In {\em 2010 48th Annual Allerton Conference on Communication,
  Control, and Computing (Allerton)}, pages 570--576. IEEE, 2010.

\bibitem{kempe2003maximizing}
David Kempe, Jon Kleinberg, and {\'E}va Tardos.
\newblock Maximizing the spread of influence through a social network.
\newblock In {\em Proceedings of the ninth ACM SIGKDD international conference
  on Knowledge discovery and data mining}, pages 137--146, 2003.

\bibitem{galam2000universality}
S~Galam.
\newblock Universality of group decision making.
\newblock In {\em Traffic and Granular Flow’99}, pages 57--67. Springer,
  2000.

\bibitem{bruinsma1984random}
R~Bruinsma.
\newblock Random-field ising model on a bethe lattice.
\newblock {\em Physical Review B}, 30(1):289, 1984.

\bibitem{nowotny2001phase}
Thomas Nowotny, Heiko Patzlaff, and Ulrich Behn.
\newblock Phase diagram of the random field ising model on the bethe lattice.
\newblock {\em Physical Review E}, 65(1):016127, 2001.

\bibitem{bleher1998phase}
Pavel~M Bleher, J~Ruiz, and Valentin~A Zagrebnov.
\newblock On the phase diagram of the random field ising model on the bethe
  lattice.
\newblock {\em Journal of statistical physics}, 93(1):33--78, 1998.

\bibitem{friedkin1997social}
Noah~E Friedkin and Eugene~C Johnsen.
\newblock Social positions in influence networks.
\newblock {\em Social networks}, 19(3):209--222, 1997.

\bibitem{yang2006mining}
Wan-Shiou Yang, Jia-Ben Dia, Hung-Chi Cheng, and Hsing-Tzu Lin.
\newblock Mining social networks for targeted advertising.
\newblock In {\em Proceedings of the 39th Annual Hawaii International
  Conference on System Sciences (HICSS'06)}, volume~6, pages 137a--137a. IEEE,
  2006.

\bibitem{stauffer2008social}
Dietrich Stauffer.
\newblock Social applications of two-dimensional ising models.
\newblock {\em American Journal of Physics}, 76(4):470--473, 2008.

\bibitem{lynn2017statistical}
Christopher~W Lynn and Daniel~D Lee.
\newblock Statistical mechanics of influence maximization with thermal noise.
\newblock {\em EPL (Europhysics Letters)}, 117(6):66001, 2017.

\bibitem{romero2020continuous}
Guillermo Romero~Moreno, Long Tran-Thanh, and Markus Brede.
\newblock Continuous influence maximisation for the voter dynamics: Is
  targeting high-degree nodes a good strategy?
\newblock In {\em Proceedings of the 19th International Conference on
  Autonomous Agents and MultiAgent Systems}, pages 1981--1983, 2020.

\bibitem{barabasi2014network}
Albert-L{\'a}szl{\'o} Barab{\'a}si.
\newblock Network science book.
\newblock {\em Network Science}, 625, 2014.

\bibitem{lynn2016maximizing}
Christopher Lynn and Daniel~D Lee.
\newblock Maximizing influence in an ising network: A mean-field optimal
  solution.
\newblock {\em Advances in neural information processing systems},
  29:2495--2503, 2016.

\bibitem{wu1982potts}
Fa-Yueh Wu.
\newblock The potts model.
\newblock {\em Reviews of modern physics}, 54(1):235, 1982.

\bibitem{blume1966theory}
M~Blume.
\newblock Theory of the first-order magnetic phase change in u o 2.
\newblock {\em Physical Review}, 141(2):517, 1966.

\bibitem{capel1966possibility}
HW~Capel.
\newblock On the possibility of first-order phase transitions in ising systems
  of triplet ions with zero-field splitting.
\newblock {\em Physica}, 32(5):966--988, 1966.

\bibitem{silvela2001breadth}
Jaime Silvela and Javier Portillo.
\newblock Breadth-first search and its application to image processing
  problems.
\newblock {\em IEEE Transactions on Image Processing}, 10(8):1194--1199, 2001.

\bibitem{brede2019transmission}
Markus Brede, Valerio Restocchi, and Sebastian Stein.
\newblock Transmission errors and influence maximization in the voter model.
\newblock {\em Journal of Statistical Mechanics: Theory and Experiment},
  2019(3):033401, 2019.

\bibitem{lynn2018maximizing}
Christopher~W Lynn and Daniel~D Lee.
\newblock Maximizing activity in ising networks via the tap approximation.
\newblock In {\em Thirty-Second AAAI Conference on Artificial Intelligence},
  2018.

\bibitem{moreno2019shielding}
Guillermo~Romero Moreno, Long Tran-Thanh, and Markus Brede.
\newblock Shielding and shadowing: A tale of two strategies for opinion control
  in the voting dynamics.
\newblock In {\em International Conference on Complex Networks and Their
  Applications}, pages 682--693. Springer, 2019.

\bibitem{leyffer2010nonlinear}
Sven Leyffer and Ashutosh Mahajan.
\newblock Nonlinear constrained optimization: methods and software.
\newblock {\em Argonee National Laboratory, Argonne, Illinois}, 60439, 2010.

\bibitem{barahona1982computational}
Francisco Barahona.
\newblock On the computational complexity of ising spin glass models.
\newblock {\em Journal of Physics A: Mathematical and General}, 15(10):3241,
  1982.

\bibitem{kotler2014measurement}
Shlomi Kotler, Nitzan Akerman, Nir Navon, Yinnon Glickman, and Roee Ozeri.
\newblock Measurement of the magnetic interaction between two bound electrons
  of two separate ions.
\newblock {\em Nature}, 510(7505):376--380, 2014.

\end{thebibliography}

\newpage
\appendix
\section*{Appendices}
\section{Detailed proof for the formula of the effective field in a junction}\label{Appendix_NonLeafProof}
In this appendix, we give the detailed calculation that proves the effective field formula for a junction,
\begin{equation}\label{Eq:EffectiveFieldJunction}
    \braket{\sigma_{k}}[h_{k}=0]=\tanh\left(\sum_{h\in \{m...n\}} \tanh^{-1}(\tanh(\beta J_{hk})\braket{\tilde\sigma_h})\right),
\end{equation}
where again $k$ is the junction spin index, and $m...n$ are the indices of the spins connected to the junction, and as before $\braket{\tilde\sigma_h}$ is the mean magnetization of the spin $h$ when the connection to the spin $k$ does not exist.

We can write the mean magnetization for $h_k=0$, using the decomposition of the Hamiltonian as in the main text, namely
\begin{eqnarray}
    \mathcal{H}(\{\sigma\}) =   \left(\sum_{i\in\{m...n\}}\mathcal{H}_i\left(\{\sigma\}\right) - J_{ki}\sigma_k\sigma_i\right) - h_k\sigma_k 
\end{eqnarray}
where 
\begin{eqnarray}
    \mathcal{H}_i(\{\sigma\}) = -\sum_{\{l,j\}\in G_i}J_{lj}\sigma_l\sigma_j - \sum_{j\in G_i}h_j\sigma_j.
\end{eqnarray}
The mean magnetization of $h_k$ is:
\begin{equation}\label{Eq:h_k}
    \braket{\sigma_k}[h_k=0]=\frac{\sum_{\{\sigma\}}\sigma_k e^{-\beta\mathcal{H}(\{\sigma\})}}{\sum_{\{\sigma\}}e^{-\beta\mathcal{H}(\{\sigma\})}}= \frac{\sum_{\{\sigma\}}\sigma_ke^{-\beta \sum_{i\in\{m...n\}}\mathcal{H}_i-J_{ki}\sigma_k\sigma_i} }{\sum_{\{\sigma\}}e^{-\beta \sum_{i\in\{m...n\}}\mathcal{H}_i-J_{ki}\sigma_k\sigma_i}}
\end{equation}
Now, we can perform the summation over $\sigma_k=\pm 1$. Let us do that first for the denominator: 
\begin{eqnarray}
    \sum_{\{\sigma\}}e^{-\beta \sum_{i\in\{m...n\}}\mathcal{H}_i-J_{ki}\sigma_k\sigma_i} &=& \sum_{\{\sigma_j\neq \sigma_k\}}e^{-\beta \sum_{i\in\{m...n\}}\mathcal{H}_i+J_{ki}\sigma_i} + \sum_{\{\sigma_j\neq \sigma_k\}}e^{-\beta \sum_{i\in\{m...n\}}\mathcal{H}_i-J_{ki}\sigma_i} \nonumber\\
    &=& \sum_{\{\sigma_j\neq \sigma_k\}}e^{-\beta \sum_{i\in\{m...n\}}\mathcal{H}_i}\cosh\left(\beta\sum_{i\in\{m...n\}}J_{ki}\sigma_i\right)
\end{eqnarray}
and similarly to the numerator:
\begin{eqnarray}
    \sum_{\{\sigma\}}\sigma_ke^{-\beta \sum_{i\in\{m...n\}}\mathcal{H}_i+J_{ki}\sigma_k\sigma_i} &=& \sum_{\{\sigma_j\neq\sigma_k\}}e^{-\beta \sum_{i\in\{m...n\}}\mathcal{H}_i+J_{ki}\sigma_i} - \sum_{\{\sigma_j\neq\sigma_k\}}e^{-\beta \sum_{i\in\{m...n\}}\mathcal{H}_i
    -J_{ki}\sigma_i} \nonumber \\
    &=&\sum_{\{\sigma_j\neq\sigma_k\}}e^{-\beta \sum_{i\in\{m...n\}}\mathcal{H}_i}\sinh\left(-\beta \sum_{i\in\{m...n\}}J_{ki}\sigma_i \right)
\end{eqnarray}
The above expressions involve hyperbolic cosine and sine of sums. We therefore next use the relations $\cos({\bf i}x) = \cosh(x)$ and $\sin({\bf i}x) = {\bf i}\sinh({\bf i}x)$ where ${\bf i}=\sqrt{-1}$, and the known identities for $\cos$ and $\sin$ of sum of angles, to write:
\begin{eqnarray}
    \cosh\left(\beta\sum_{i\in\{m...n\}}J_{ki}\sigma_i\right) = \cos\left({\bf i}\beta\sum_{i\in\{m...n\}}J_{ki}\sigma_i\right) = \sum_{r \hbox{  even}}(-1)^{r/2}\sum_{A\subseteq\{1,2,...\},|A|=r}\left(\prod_{h\in A}\sin({\bf i}\beta J_{kh}\sigma_h) \prod_{h\not\in A}\cos({\bf i}\beta J_{kh}\sigma_h) \right)\nonumber
    \end{eqnarray}
    \begin{eqnarray}
    &=&\sum_{r \hbox{ even}}(-1)^{r/2}\sum_{A\subseteq\{m...n\},|A|=r}{\bf i}^{r}\left(\prod_{h\in A}\sinh(\beta J_{kh}\sigma_h) \prod_{h\not\in A}\cosh(\beta J_{kh}\sigma_h) \right)\nonumber\\
    &=&\sum_{A\subseteq\{m...n\},|A| \hbox{ even}}\left(\prod_{h\in A}\sigma_h\sinh(\beta J_{kh}) \prod_{h\not\in A}\cosh(\beta J_{kh}) \right)
\end{eqnarray}
where we have used ${\bf i}^r = (-1)^{r/2}$, and in the last line that $\cosh(\beta J_{kh}) = \cosh(\beta J_{kh}\sigma_h)$ and $\sinh(\beta J_{kh}\sigma_h)=\sigma_h\sinh(\beta J_{kh})$for both $\sigma_h=1$ and $\sigma_h=-1$.
A similar expression for $\sinh$ of a sum is given by:
\begin{eqnarray}
    \sinh\left(-\beta \sum_{i\in\{m...n\}}J_{ki}\sigma_i \right) &=& -{\bf i}\sinh\left(-{\bf i}\beta \sum_{i\in\{m...n\}}J_{ki}\sigma_i \right) \nonumber\\
    &=&-{\bf i}\sum_{r\hbox{ odd}} (-1)^{(r-1)/2} \sum_{A\subseteq\{m...n\},|A|=r}\left(\prod_{h\in A}\sin(-{\bf i}\beta J_{kh}\sigma_h) \prod_{h\not\in A}\cos(-{\bf i}\beta J_{kh}\sigma_h) \right)\nonumber\\
    &=&{\bf i}\sum_{r\hbox{ odd}} (-1)^{(r-1)/2} \sum_{A\subseteq\{m...n\},|A|=r}\left(\prod_{h\in A}{\bf i}\sinh(\beta J_{kh}\sigma_h) \prod_{h\not\in A}\cosh(\beta J_{kh}\sigma_h) \right)\nonumber\\
    &=&\sum_{r\hbox{ odd}} (-1)^{(r-1)/2} \sum_{A\subseteq\{m...n\},|A|=r}{\bf i}^{r+1}\left(\prod_{h\in A}\sinh(\beta J_{kh}\sigma_h) \prod_{h\not\in A}\cosh(\beta J_{kh}\sigma_h) \right)\nonumber\\
    &=&-\sum_{A\subseteq\{m...n\},|A|\hbox{ odd}}\left(\prod_{h\in A}\sigma_h\sinh(\beta J_{kh}) \prod_{h\not\in A}\cosh(\beta J_{kh}) \right)
\end{eqnarray}
Using the above in Eq.(\ref{Eq:h_k}), we can write
\begin{eqnarray}\label{Eq:Sigma_h_k_derive_1}
    \braket{\sigma_k}[h_k=0] &=& \frac{\sum_{\{\sigma_j\neq \sigma_k\}}\sum_{A\subseteq\{m...n\},|A| \hbox{ odd}}\left(\prod_{h\in A}\sigma_h\sinh(\beta J_{kh})e^{-\beta \mathcal{H}_h} \prod_{h\not\in A}\cosh(\beta J_{kh})e^{-\beta \mathcal{H}_h} \right)}{\sum_{\{\sigma_j\neq \sigma_k\}}\sum_{A\subseteq\{m...n\},|A| \hbox{ even}}\left(\prod_{h\in A}\sigma_h\sinh(\beta J_{kh})e^{-\beta \mathcal{H}_h} \prod_{h\not\in A}\cosh(\beta J_{kh})e^{-\beta \mathcal{H}_h} \right)}\nonumber\\
    &=&\frac{\sum_{A\subseteq\{m...n\},|A| \hbox{ odd}}\left(\prod_{h\in A}\braket{\tilde \sigma_h}\sinh(\beta J_{kh})\tilde Z_{h} \prod_{h\not\in A}\cosh(\beta J_{kh}) \tilde Z_{h}\right)}{\sum_{A\subseteq\{m...n\},|A| \hbox{ even}}\left(\prod_{h\in A}\braket{\tilde \sigma_h}\tilde Z_{h}\sinh(\beta J_{kh}) \prod_{h\not\in A}\cosh(\beta J_{kh})\tilde Z_{h}\right)}\nonumber\\
    &=&\frac{\left(\prod_{h\in\{m...n\}}\tilde Z_h\right)\sum_{A\subseteq\{m...n\},|A| \hbox{ odd}}\left(\prod_{h\in A}\braket{\tilde \sigma_h}\sinh(\beta J_{kh}) \prod_{h\not\in A}\cosh(\beta J_{kh})\right)}{\left(\prod_{h\in\{m...n\}}\tilde Z_h\right)\sum_{A\subseteq\{m...n\},|A| \hbox{ even}}\left(\prod_{h\in A}\braket{\tilde \sigma_h}\sinh(\beta J_{kh}) \prod_{h\not\in A}\cosh(\beta J_{kh})\right)}\nonumber\\
    &=&\frac{\sum_{A\subseteq\{m...n\},|A| \hbox{ odd}}\left(\prod_{h\in A}\braket{\tilde \sigma_h}\sinh(\beta J_{kh}) \prod_{h\not\in A}\cosh(\beta J_{kh})\right)}{\sum_{A\subseteq\{m...n\},|A| \hbox{ even}}\left(\prod_{h\in A}\braket{\tilde \sigma_h}\sinh(\beta J_{kh}) \prod_{h\not\in A}\cosh(\beta J_{kh})\right)}
\end{eqnarray}
where $\tilde Z_{h}$ is the partition function associated with $\mathcal{H}_h$.

Next, we show that this expression is also equal to 
\begin{equation}
    \braket{\sigma_{k}}[h_{k}=0]=\tanh\left(\sum_{h\in \{m...n\}} \tanh^{-1}(\tanh(\beta J_{hk})\braket{\tilde\sigma_h})\right)
\end{equation}
Let us open this expression using the formula for $\tanh\left(\sum \theta_i \right)$, which can be derived from those of $\sinh$ and $\cosh$ of sums we used above. The general formula is given by
\begin{equation}
    \tanh\left(\sum \theta_i \right)= \frac{\sum_{\text{odd r}}\sum_{A\subseteq\{1,2,3...\},|A|=r}\prod_{i\in A}\tanh(\theta_{i})}{\sum_{\text{even r}\ge0}\sum_{A\subseteq\{1,2,3...\},|A|=r}\prod_{i\in A}\tanh(\theta_{i})}
\end{equation}
and applying it to our case gives:
\begin{eqnarray}
    \tanh\left(\sum_{h\in \{m...n\}} \tanh^{-1}(\tanh(\beta J_{hk})\braket{\tilde\sigma_h})\right) = \frac{\sum_{A\subseteq\{m...n\},|A|=r \hbox{ odd}}\prod_{h\in A}\tanh(\beta J_{hk})\braket{\tilde\sigma_h}}{\sum_{A\subseteq\{m...n\},|A|=r\hbox{ even}}\prod_{h\in A}\tanh(\beta J_{hk})\braket{\tilde\sigma_h}}\nonumber\\
    =\frac{\sum_{A\subseteq\{m...n\},|A|=r \hbox{ odd}}\prod_{h\in A}\frac{\sinh(\beta J_{hk})\braket{\tilde\sigma_h}}{\cosh(\beta J_{hk})}}{\sum_{A\subseteq\{m...n\},|A|=r\hbox{ even}}\prod_{h\in A}\frac{\sinh(\beta J_{hk})\braket{\tilde\sigma_h}}{\cosh(\beta J_{hk})}}\nonumber\\
    =\frac{\sum_{A\subseteq\{m...n\},|A|=r \hbox{ odd}}\left(\prod_{h\in\{m...n\}}\cosh^{-1}(\beta J_{hk})\right)\prod_{h\in A}\sinh(\beta J_{hk})\braket{\tilde\sigma_h}\prod_{h\not\in A}\cosh(\beta J_{hk})}{\sum_{A\subseteq\{m...n\},|A|=r\hbox{ even}}\left(\prod_{h\in\{m...n\}}\cosh^{-1}(\beta J_{hk})\right)\prod_{h\in A}\sinh(\beta J_{hk})\braket{\tilde\sigma_h}\prod_{h\not\in A}\cosh(\beta J_{hk})}\nonumber\\
    =\frac{\sum_{A\subseteq\{m...n\},|A|=r \hbox{ odd}}\prod_{h\in A}\sinh(\beta J_{hk})\braket{\tilde\sigma_h}\prod_{h\not\in A}\cosh(\beta J_{hk})}{\sum_{A\subseteq\{m...n\},|A|=r\hbox{ even}}\prod_{h\in A}\sinh(\beta J_{hk})\braket{\tilde\sigma_h}\prod_{h\not\in A}\cosh(\beta J_{hk})}
\end{eqnarray}
which is exactly the end result of Eq.(\ref{Eq:Sigma_h_k_derive_1}). We therefore proved Eq.(\ref{Eq:EffectiveFieldJunction}), as required.


\section{Pseudo-code for the algorithm that finds the mean magnetization}\label{Sec:AppAlgo}

\begin{algorithm}[H]
\DontPrintSemicolon
  
  \KwInput{Graph,field\_array}
  \KwOutput{total\_mag}
  
  \textit{total\_mag} = 0
          \\ \For{\textit{node} $\in$ \textit{Graph}}   
        { 
         BFS(\textit{Graph,node})
\\ \textit{total\_mag}+=$\tanh(node.h+Recursive\_Effective\_Field(\textit{node}))$
        }

    \Return  \textit{total\_mag}
  
\caption{Network Total Magnetization}
\end{algorithm}

\begin{algorithm}[H]
\DontPrintSemicolon
  
  \KwInput{root} 
  \KwOutput{node\_effective\_field}

    \textit{node\_mag} = \textit{root}.h
   
        \For{\textit{children} in \textit{root}.children}    
        { 
        	 h=Recursive\_Effective\_Field(children.h)
        	 \\ \textit{node\_mag} +=$ \tanh^{-1}(\tanh(\beta J) \tanh(\beta h ))$
        }

    \Return \textit{node\_effective\_field}
  
\caption{Recursive\_Effective\_Field}
\end{algorithm}

\begin{algorithm}[H]
\DontPrintSemicolon
  
  \KwInput{Graph,root}
\textit{nodes\_to\_visit} = []
\\  \textit{visited\_nodes}  = []
 \textit{nodes\_to\_visit.add\_element(root})
\\ \textit{visited\_nodes.add\_element(root})
\\\While{\textit{nodes\_to\_visit} is not empty}
    { \textit{parent} =
	\textit{nodes\_to\_visit}.first\_element()
	\\ \textit{visited\_nodes}.add\_element(\textit{parent})
	\\ \textit{nodes\_to\_visit}.delete\_first\_element()
	\\ \For{\textit{neighbor} in \textit{parent}.neighbours()}
	{
	  \If{ \textit{neighbor} $\not\in$  \textit{visited\_nodes}}
{	 
\textit{parent.children.add\_element(neighbor)}}
	 
 \textit{neighbor.parent} = \textit{parent}
\\ \textit{nodes\_to\_visit.add\_element(neighbor)}
 }}
 \Return \textit{visited\_nodes}
 \caption{BFS algorithm}
\end{algorithm}

\end{document}